\documentstyle[12pt,epsf,epsfig,cite,rotating]{article}

  \def\thebibliography#1{\section*{References}\list
   {[\arabic{enumi}]}{\settowidth\labelwidth{[#1]}\leftmargin\labelwidth
   \advance\leftmargin\labelsep
   \usecounter{enumi}}
   \def\newblock{\hskip .11em plus .33em minus -.07em}
   \sloppy
   \sfcode`\.=1000\relax}
  
  \newcounter{cap}
  {\begin{list}{Figure \arabic{cap}\hfil}{\usecounter{cap}
  \settowidth{\labelwidth}{Figure #1}%
  \setlength{\leftmargin}{\labelwidth}%
  \addtolength{\leftmargin}{\labelsep}%
  \setlength{\parsep}{2mm plus 1mm minus 1mm}
  \setlength{\itemsep}{3mm plus 2mm minus 2mm}
  }}%
  {\end{list}}
  {\begin{list}{}{\settowidth{\labelwidth}{#1}%
  \setlength{\leftmargin}{\labelwidth}%
  \addtolength{\leftmargin}{\labelsep}%
  \setlength{\itemsep}{0pt plus 1pt}
  \setlength{\parsep}{0pt plus 1pt}
  \setlength{\topsep}{0pt plus 1pt}
  \setlength{\partopsep}{0pt plus 1pt}
  \setlength{\parskip}{2mm plus 1mm minus 1mm}
  }}%
  {\end{list}}

\newcommand{\gapprox}{\stackrel{>}{_{\sim}}}
\newcommand{\lapprox}{\stackrel{<}{_{\sim}}}
\newcommand{\pom}{\rm I\!P}
\newcommand{\reg}{\rm I\!R}
\newcommand{\pt}{p_{_T}}
\newcommand{\ptj}{p_{_T}^{\rm jet}}
\newcommand{\etj}{E_{_T}^{\rm jet}}

\newcommand{\mx}{M_{_X}}
\newcommand{\my}{M_{_Y}}
\newcommand{\alphapom}{\alpha_{_{\rm I\!P}}}
\newcommand{\alphareg}{\alpha_{_{\rm I\!R}}}

\newcommand{\zpom}{z_{_{\rm \pom}}}
\newcommand{\xgam}{x_{\gamma}}
\newcommand{\zpomj}{z_{_{\rm \pom}}^{\rm jets}}
\newcommand{\xgamj}{x_{\gamma}^{\rm jets}}
\newcommand{\scaption}[1]{\caption{\protect{\footnotesize  #1}}}
\newcommand{\xpom}{x_{_{\pom}}}

\newcommand{\shatj}{\hat{s}^{\rm jets}}
\newcommand{\pthat}{\hat{p}_{_T}}
\newcommand{\pthatsq}{\hat{p}_{_{T}}^{\hspace{0.05cm} 2}}
\def\etalk{{ et al., }}

\begin{document}

\begin{titlepage}
\begin{flushleft}
DESY 98--092 \hspace{10cm} ISSN 0418-9833 \\
August 1998
\end{flushleft}
\vspace*{3.cm}
\begin{center}
  \begin{Large}
   \boldmath \bf{Diffractive dijet production at HERA \\}
   \unboldmath

  \vspace*{2.cm} H1 Collaboration \\ 
\end{Large}

\vspace*{1cm}

\end{center}

\vspace*{1cm}

\begin{abstract}
\noindent
Interactions of the type 
$ep \rightarrow e X Y$ are studied, where
the component $X$ of the hadronic final state
contains two jets and is
well separated in rapidity from a leading baryonic
system $Y$. Analyses are performed of both resolved and direct 
photoproduction and of 
deep-inelastic scattering with photon virtualities in the
range $7.5 < Q^2 < 80 \ {\rm GeV^2}$. 
Cross sections are presented where
$Y$ has mass $\my < 1.6 \ {\rm GeV}$, the
squared four-momentum transferred at the proton vertex satisfies
$|t| < 1 \ {\rm GeV^2}$ and the two jets  each have transverse
momentum $\ptj > 5 \ {\rm GeV}$ relative to the photon direction
in the rest frame of $X$. 
Models based on a factorisable diffractive
exchange with a gluon dominated 
structure, evolved to a scale set by the transverse
momentum $\pthat$ of the outgoing partons
from the hard interaction, 
give good descriptions of the data. 
Exclusive $q \bar{q}$ production, as calculated in perturbative QCD using the 
squared proton gluon density, represents at most a
small fraction of the measured cross section.
The compatibility of the data with a breaking of diffractive factorisation
due to spectator interactions in resolved photoproduction is investigated.
\end{abstract}
\end{titlepage}


\vfill
\clearpage
\begin{sloppypar}
\noindent
 C.~Adloff$^{34}$,                
 M.~Anderson$^{21}$,              
 V.~Andreev$^{24}$,               
 B.~Andrieu$^{27}$,               
 V.~Arkadov$^{34}$,               
 C.~Arndt$^{10}$,                 
 I.~Ayyaz$^{28}$,                 
 A.~Babaev$^{23}$,                
 J.~B\"ahr$^{34}$,                
 J.~B\'an$^{16}$
% ...
 \\
\bigskip 

\noindent
{\footnotesize{
% ...
$ ^{10}$ DESY, Hamburg, Germany$^a$ \\
% ...
$ ^{16}$ Institute of Experimental Physics, Slovak Academy of
          Sciences, Ko\v{s}ice, Slovak Republic$^{f,j}$ \\
% ...
$ ^{21}$ Department of Physics and Astronomy, 
          University of Manchester, Manchester, UK$^b$ \\
% ...
$ ^{23}$ Institute for Theoretical and Experimental Physics,
          Moscow, Russia \\
 $ ^{24}$ Lebedev Physical Institute, Moscow, Russia$^{f,k}$ \\
% ...
$ ^{27}$ LPNHE, Ecole Polytechnique, IN2P3-CNRS, Palaiseau, France \\
 $ ^{28}$ LPNHE, Universit\'{e}s Paris VI and VII, IN2P3-CNRS,
          Paris, France \\
% ...
$ ^{34}$ DESY, Institut f\"ur Hochenergiephysik, Zeuthen, Germany$^a$ \\
% ...
$ ^a$ Supported by the Bundesministerium f\"ur Bildung, Wissenschaft,
        Forschung und Technologie, FRG,
        under contract numbers 7AC17P, 7AC47P, 7DO55P, 7HH17I, 7HH27P,
        7HD17P, 7HD27P, 7KI17I, 6MP17I and 7WT87P \\
 $ ^b$ Supported by the UK Particle Physics and Astronomy Research
       Council, and formerly by the UK Science and Engineering Research
       Council \\
% ...
$ ^f$ Supported by the Deutsche Forschungsgemeinschaft \\
% ...
$ ^j$ Supported by VEGA SR grant no. 2/1325/96 \\
 $ ^k$ Supported by Russian Foundation for Basic Research 
       grant no. 96-02-00019 
}}

\end{sloppypar}


\section{Introduction}

The concept of a pomeron ($\pom$) trajectory possessing vacuum quantum 
numbers and mediating diffractive scattering has proved remarkably 
successful in formulating a Regge description of high energy hadronic cross 
sections\cite{diff:review}. 
There has been considerable recent 
interest in understanding the underlying dynamics of diffractive 
interactions in terms of quantum chromodynamics (QCD). 
If the pomeron can be
considered as a partonic system\cite{diff:hardscat}, then perturbatively 
calculable processes involving large transverse momenta are expected.
High transverse momentum jet production has been observed in diffractive
$p\bar{p}$ scattering \cite{UA8:hard,UA8:super,tevatron:jets} and also in
photoproduction at HERA\cite{H1:gpdiffjet,ZEUS:gpdiffjet,ZEUS:jetnew}. 

At HERA, diffractive scattering is 
studied both in photoproduction and at large 
$Q^2$ using events of the type $ep \rightarrow eXY$, where the 
hadronic systems $X$ and $Y$ are separated by a large region of 
pseudorapidity that is devoid of hadronic
activity \cite{ZEUS:difobs,H1:difobs} and $Y$ is predominantly a 
proton \cite{ZEUS:LPS}. 
The contribution from such processes to the total 
photoproduction cross section at $\gamma p$ centre of mass energies 
$W \sim 200 \ {\rm GeV}$ exceeds 20\% \cite{H1:gammap,ZEUS:gammap}.
The leading twist large rapidity gap component
represents approximately 10\% of the total deep-inelastic scattering (DIS)
cross section \cite{F2D393,ZeusF2D3,F2D394}.

The contribution to the proton structure function from the process 
$ep \rightarrow eXY$ \linebreak
($\my < 1.6 \ {\rm GeV}$, $|t| < 1 \ {\rm GeV^2}$)
has been measured differentially in the fraction $\xpom$ of the proton beam 
momentum transferred to the system $X$. The results \cite{F2D394}
have been presented
in terms of a structure function 
$F_{2}^{D(3)}(\xpom,\beta,Q^2)$ where
$\beta = x / \xpom$ and $x$ is the Bjorken scaling variable. 
A Regge analysis of 
$F_2^{D(3)}$
demonstrates that diffraction is dominant at
small $\xpom$, with sub-leading exchanges ($\reg$) becoming
important as $\xpom$ increases. 
The $\beta$ and $Q^2$ dependence of $F_2^{D(3)}$ 
has been considered in terms of the QCD evolution
of a structure function for the pomeron \cite{F2D393,F2D394}. 
Assuming that the evolution for $\beta < 0.65$ is 
governed by the DGLAP\cite{DGLAP} 
equations, diffractive parton 
distributions are extracted that are dominated 
at low $Q^2$ by gluons carrying large fractions of the
exchanged momentum (referred to as `hard' gluons in this paper).
Under the hypothesis of diffractive factorisation,
the parton distributions for the pomeron extracted from $F_2^{D(3)}$ 
are expected
to describe diffractive interactions wherever perturbation theory
may be applied. 

Complementary to the approach based on pomeron parton distributions,
diffractive $\gamma^* p$ interactions have also been modelled in
terms of the elastic
scattering from the proton of partonic fluctuations of the 
photon, which at high energy develop well in advance of the target. 
The kinematic dependences expected for the distinct photon fluctuations
are very different. 
For the simplest $q \bar{q}$ state, the colour
transparency mechanism suppresses the cross section when the transverse 
separation of the two partons is small \cite{AJM:ct}. The $q \bar{q}$ cross
section is therefore expected to be 
significant only at comparatively small 
transverse momenta $\pthat$
of the outgoing partons in the centre of mass frame of
the hard interaction. In this case, the entire diffractive mass 
$\mx$ is shared by the $q \bar{q}$ pair. 
Several calculations \cite{gg:proton,qqg,gg:rapgap} have been 
performed, the scattering of the $q \bar{q}$ pair from the proton 
being modelled
by the exchange of two gluons in a net colour singlet
configuration \cite{low:nussinov}. In contrast to the $q \bar{q}$ state, the
colour transparency effect is prevented for photon fluctuations with 
additional low transverse momentum gluons.
At large $\pthat$ the cross section is therefore 
expected to be dominated by the scattering of
$q \bar{q} g$ and higher multiplicity components of the photon.
This conclusion is reached both in two gluon exchange models \cite{qqg}
and in 
a semi-classical model \cite{buchmuller:1996} in which the partonic photon 
fluctuations are scattered in the proton colour field \cite{buchmuller:1997}.
A recent parameterisation in 
terms of contributions from $q \bar{q}$ and $q \bar{q} g$ states can
give an acceptable description of $F_2^{D(3)}$ data \cite{bartels:fullmodel}.

Several hadronic final state observables 
are sensitive to the partonic structure of diffractive interactions. 
The parton distributions extracted from the QCD analysis of 
$F_2^{D(3)}$ are able to describe measurements of 
event thrust\cite{H1:thrust}, 
energy flow, charged particle spectra\cite{H1:eflow}
and charged particle multiplicities and their correlations\cite{H1:multip}
in diffractive DIS. Viewed in terms of photon fluctuations, 
these measurements
confirm the need for Fock states with one or more gluons
at comparatively large $\mx$ and high $\pthat$.  

In this paper, large rapidity gap events that contain two 
high transverse momentum jets
as components of the system $X$ are investigated in 
separate analyses of photoproduction and DIS data. 
Cross section measurements are presented 
differentially in the jet pseudorapidity in the laboratory frame, in the jet 
momentum transverse to the
$\gamma^{(*)}$ axis in the rest frame of the system $X$ and
in the fractions of the photon and pomeron 
momenta that are transferred to the dijet system. 
The data are 
compared to predictions based on diffractive
parton densities extracted from $F_2^{D(3)}$ at a scale set by
$\pthat$. 
In diffractive interactions where 
both colliding particles can interact strongly,
it has been argued that the factorisation property for 
diffractive parton distributions 
is at best approximate, since
additional soft interactions among spectator partons may couple
the extended hadrons to one another \cite{survive,glm:survive,fac:break}.
The possible presence of such an effect in resolved photoproduction, 
where the photon interacts through its hadronic structure,
is investigated.
A model of the contribution to the diffractive DIS cross section from the 
$q \bar{q}$ fluctuation of the photon \cite{gg:rapgap}
is also compared to the data. A further comparison is made with an inclusive
model of DIS in which soft interactions produce
rapidity gaps by altering the colour connections 
between outgoing partons \cite{sci}.

In section~\ref{sigmadef}, kinematic variables are 
introduced and the measured processes
are defined at the level of final state hadrons. Section~\ref{mcmods} 
introduces the Monte Carlo models which are used in the procedure to correct
the data for experimental bias and finite acceptance.
The components of the H1 detector most relevant to the analysis, the event 
selection and reconstruction 
methods are covered in section~\ref{procedure}, along with further details 
of the experimental procedure. The measured
cross sections are presented and discussed in section~\ref{results}.


\section{Kinematic variable and cross section definitions}
\label{sigmadef}

\begin{figure}[htb]
 \begin{center}
   \begin{picture}(160,120)
     \put(-120,0){\epsfig{file=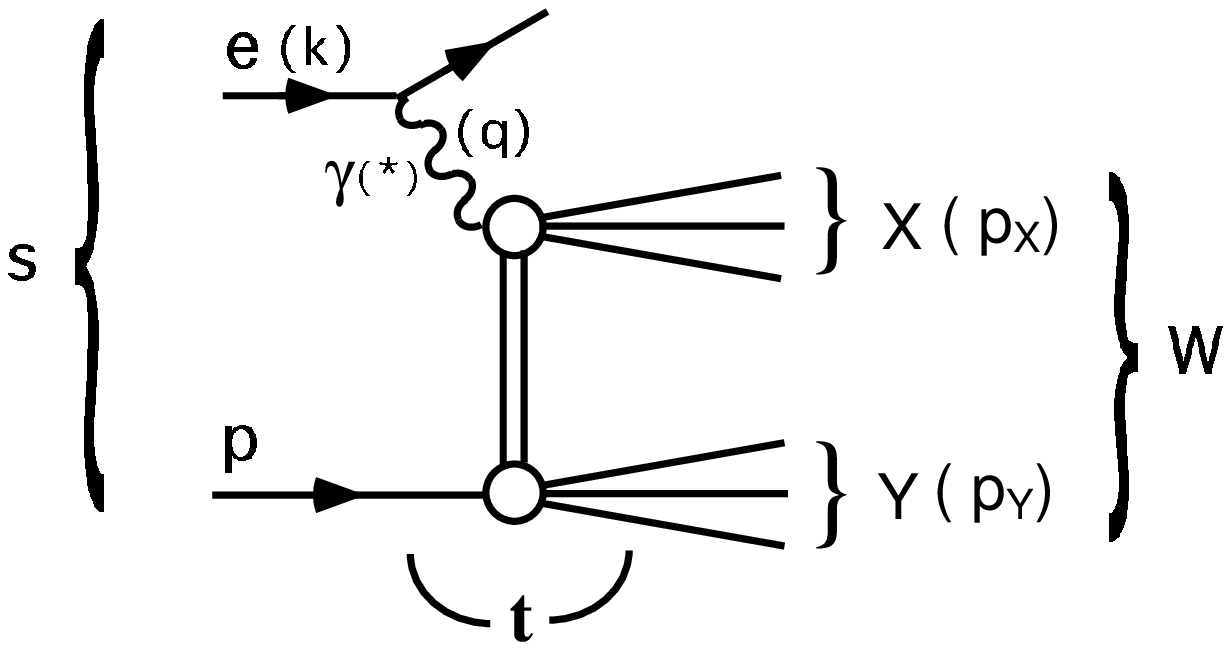,width=0.4\textwidth}}
     \put(120,2){\epsfig{file=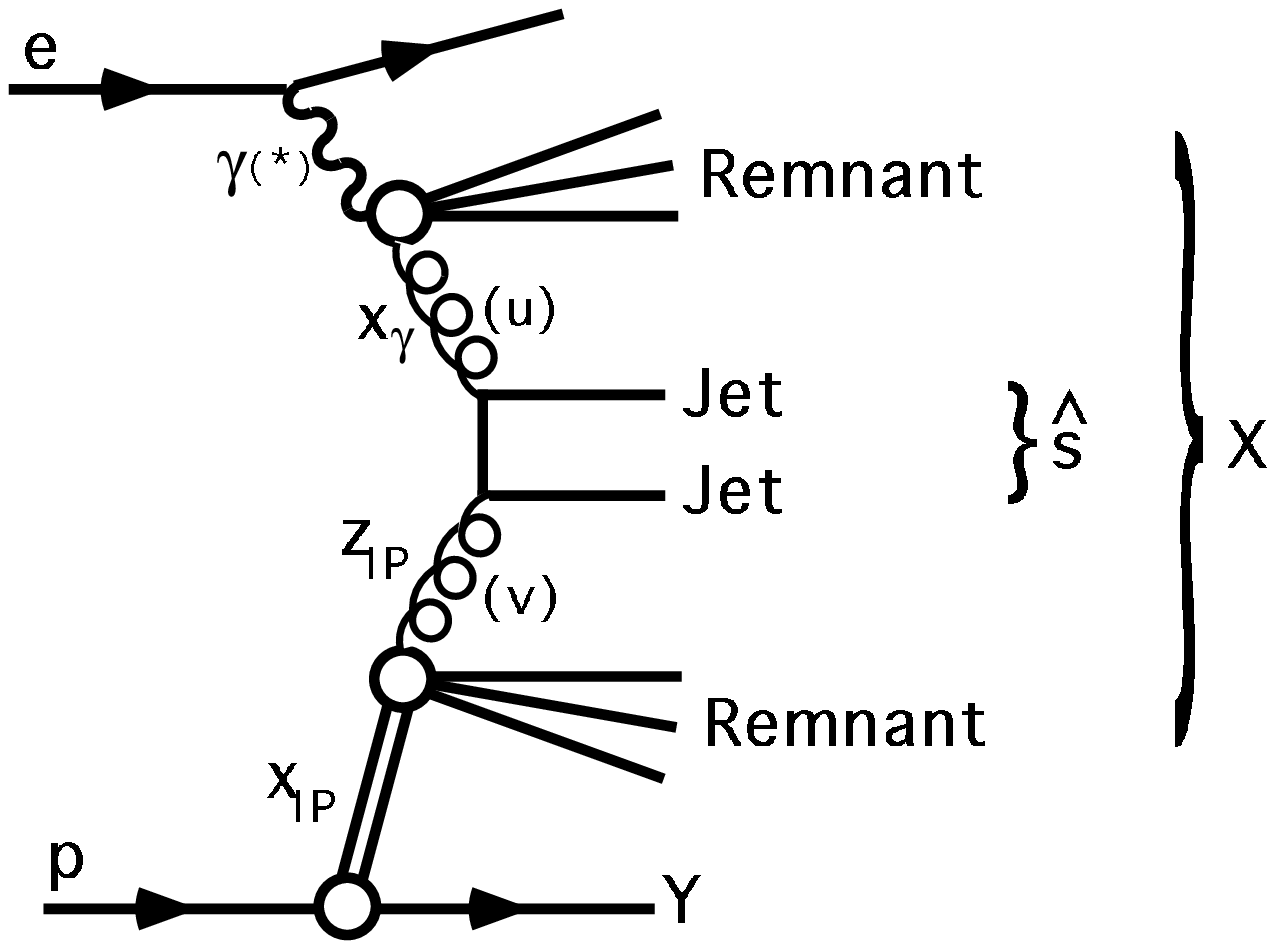,width=0.4\textwidth}}
     \put(-60,-15){\bf{(a)}}
     \put(150,-15){\bf{(b)}}
   \end{picture}
 \end{center}
  \scaption  {(a) Illustration of the generic process, $ep \rightarrow eXY$,
in which the largest gap in the rapidity distribution of final state hadrons
separates the systems $X$ and $Y$, associated with
the photon and proton vertices respectively. 
(b) An example diffractive
dijet production process in which the photon is resolved. 
The partons entering the 
hard interaction from the photon and the pomeron 
have 4-momenta $u$ and $v$ and carry momentum fractions
$\xgam$ and $\zpom$ respectively.}
  \label{basic}
\end{figure}

The hadronic final state is
considered here in terms of the generic quasi-two body photon-proton 
interaction $\gamma^{(*)} p \rightarrow XY$,
as illustrated in figure~\ref{basic}a. By definition, the systems $X$ and $Y$ 
are separated by the largest gap in the rapidity distribution of final state
hadrons and $Y$ is the system closest to the outgoing proton 
direction \cite{H1:gammap,F2D394}.

With $k$ and $P$ denoting the 4-vectors of the incoming electron and proton
respectively and $q$ the 4-vector of the photon, the standard
kinematic variables
\begin{eqnarray}
s \equiv (k + P)^2 \hspace{1.5cm} Q^2 \equiv -q^2 
\hspace{1.5cm} W^2 \equiv (q + P)^2
\hspace{1.5cm} y \equiv \frac{q \cdot P}{k \cdot P}
\end{eqnarray}
are defined. With $p_{_X}$ and $p_{_Y}$ representing the 
4-vectors of the two 
distinct components of the hadronic final 
state in the context of figure~\ref{basic}a, 
the data are also discussed in terms of
\begin{eqnarray}
\mx^2 \equiv p_{_X}^2 \hspace{1.5cm} \my^2 \equiv p_{_Y}^2 \hspace{1.5cm} 
t \equiv (P - p_{_Y})^2  
\hspace{1.5cm} \xpom \equiv \frac{q \cdot (P - p_{_Y})}{q \cdot P} \ ,
\end{eqnarray}
where $\mx$ and $\my$ are the invariant masses of $X$ and $Y$,
$t$ is the squared four-momentum transferred between the photon and the 
incoming proton and $\xpom$ is the fraction of the proton beam
momentum transferred to the system $X$. 
Further kinematic variables are defined for use in the large $Q^2$ regime:
\begin{eqnarray}
x \equiv \frac{Q^2}{2 q \cdot P} \hspace{1.5cm} 
\beta \equiv \frac{x}{\xpom} \equiv \frac{Q^2}{2 q \cdot (P - p_{_Y})} \ ,
\label{betaeq}
\end{eqnarray}
where $\beta$ may be
interpreted in the proton infinite momentum frame as the 
fraction of the exchanged 4-momentum that is carried by the
quark coupling to the photon. 

Further variables describing the hard interaction are introduced for the 
case of dijet production. Assuming a ``2 $\rightarrow$ 2 parton'' 
interaction of the
kind shown in figure~\ref{basic}b, the system $X$ generally contains low
transverse momentum remnants of the photon and of the pomeron. If the partons 
from the photon and the pomeron 
entering the hard scattering 
have 4-momenta $u$ and $v$ respectively, the 
dijet system has squared invariant mass
\begin{eqnarray}
  \hat{s} \equiv (u + v)^2 \ ,
\end{eqnarray}
and the projections
\begin{eqnarray}
  \xgam \equiv \frac{P \cdot u}{P \cdot q} \hspace{1.5cm} 
  \zpom \equiv \frac{q \cdot v}{q \cdot (P - p_{_Y})}
\label{zpeq}
\end{eqnarray}
yield the fractions of the photon and pomeron momenta respectively
carried by the partons involved in the hard interaction.

Hadron level cross sections are measured differentially in the 
estimators\footnote{Throughout this paper, 
hadron level variables are
represented with the superscripts `jet' or `jets' in 
order to distinguish them from parton level quantities.}
$\xgamj$ and $\zpomj$ of the variables $\xgam$ and 
$\zpom$. 
In the photoproduction analysis, the jet variables are defined as
\begin{eqnarray}
\xgamj \equiv 
\frac{\sum_{\rm jets} (E_i - p_{z,i})}{\sum_{\rm X} (E_i-p_{z,i})} 
\hspace{1.5cm}
\zpomj \equiv
\frac{\sum_{\rm jets} (E_i + p_{z,i})}{\sum_{\rm X} (E_i+p_{z,i})} 
\ ,
\label{shatetc1}
\end{eqnarray}
where the sums labelled `jets' and `X' run over all hadrons
attributed to the dijet system and to the full system $X$ respectively.
The hadron energies $E_i$ and longitudinal momenta $p_{z,i}$ are calculated
in the HERA laboratory frame, the positive $z$ direction being that of the
proton beam. In DIS, the 
relationship\footnote{This relationship is derived from 
equations (~\ref{betaeq} -~\ref{zpeq})
assuming that the photon interacts directly ($u = q$)
and that the parton $v$ entering the hard
scattering is massless.} 
\begin{eqnarray}
\zpomj \equiv \beta (1 + \frac{\shatj}{Q^2}) 
\label{shatetc2}
\end{eqnarray}
defines the hadron level variable, where
\begin{eqnarray}
\shatj \equiv ( \sum_{\rm jets} E_i )^{2} - 
( \sum_{\rm jets} \vec{p}_i )^2 \ .
\label{shatetc3}
\end{eqnarray}

\section{Monte Carlo simulations}
\label{mcmods}

Monte Carlo simulations are used to correct the data for detector 
inefficiencies and for migrations of kinematic quantities due to the finite 
resolution of the reconstruction. For all events generated, the H1 detector 
response 
is simulated in detail and the Monte Carlo events are subjected to the
same analysis chain as the data. 

Hard diffractive scattering in photoproduction is modelled using the
POMPYT 2.6 \cite{POMPYT} simulation, which is a 
diffraction-specific 
extension to PYTHIA\cite{PYTHIA}, containing \linebreak
both direct ($\xgam = 1$) and
resolved ($\xgam < 1$) photon interactions. For DIS,
the 
RAPGAP 2.02 \cite{RAPGAP} model is used. In both cases,
a partonic sub-structure is ascribed to the pomeron. The diffractive
contribution $\sigma^{\rm D}$ to the cross section takes the form
\begin{eqnarray}
{\rm d} \sigma^{\rm D}(e p \rightarrow e p X) =
f_{\pom / p}(\xpom,t) \cdot
{\rm d} \sigma^{e \pom \rightarrow e X} (\mu, \xgam, \zpom) \ ,
\label{difxsec}
\end{eqnarray}
where $f_{\pom / p}(\xpom,t)$ represents the pomeron flux
associated with the beam proton and 
${\rm d} \sigma^{e \pom \rightarrow e X} (\mu, \xgam, \zpom)$ 
is the cross section for the electron-pomeron hard
interaction at a scale $\mu$. In POMPYT, a flux $f_{\gamma/e}(y,Q^2)$
of transverse photons is
factorised from the beam electron using the equivalent photon approximation
\cite{ww:etc}, such that
\begin{eqnarray}
{\rm d} \sigma^{e \pom \rightarrow e X} (\mu, \xgam, \zpom) =
f_{\gamma/e}(y,Q^2) \cdot
{\rm d} \sigma^{\gamma \pom \rightarrow X} (\mu, \xgam, \zpom) \ .
\end{eqnarray}

In both models, the pomeron flux factor is taken to be
\begin{eqnarray}
  f_{\pom / p}(\xpom,t) = \left( \frac{1}{\xpom}
  \right)^{2 \alphapom (t) - 1} e^{b_{_{\pom}} t} \ ,
\label{flux}
\end{eqnarray}
with trajectory $\alphapom(t) = 1.20 + 0.26 \, t$ and slope parameter 
$b_{_{\pom}} = 4.6 \ {\rm GeV^{-2}}$.
This $\xpom$ dependence matches that extracted by H1 in a fit to $F_2^{D(3)}$
(fit B of \cite{F2D394}), the normalisation and $t$ dependences
being the same as those assumed in that fit.\footnote{The 
normalisations of the pomeron flux factor [equation (~\ref{flux})] and the
$e \pom$ cross section are separately ambiguous, though their product
[equation (~\ref{difxsec})] is constrained by the measurement of
$F_2^{D(3)}$.}
The $t$ dependence is consistent with that recently measured in 
diffractive DIS by the ZEUS collaboration \cite{ZEUS:LPS}.

\begin{figure}[h] \unitlength 1mm
 \begin{center}
   \begin{picture}(100,30)
     \put(60,0){\epsfig{file=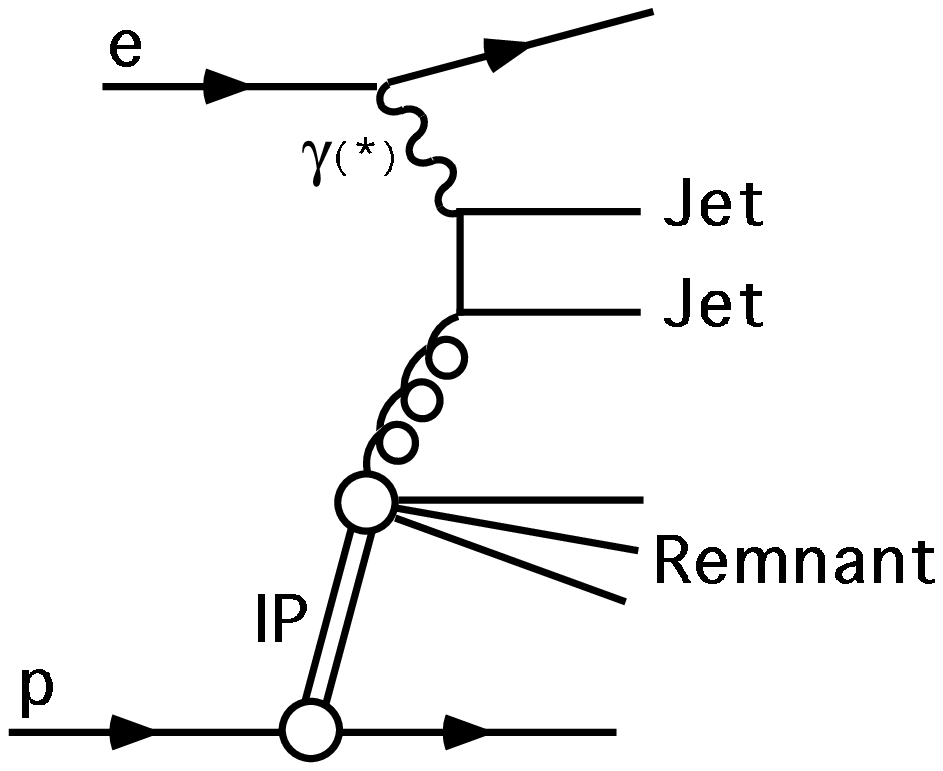,width=0.28\textwidth}}
     \put(-10,2){\epsfig{file=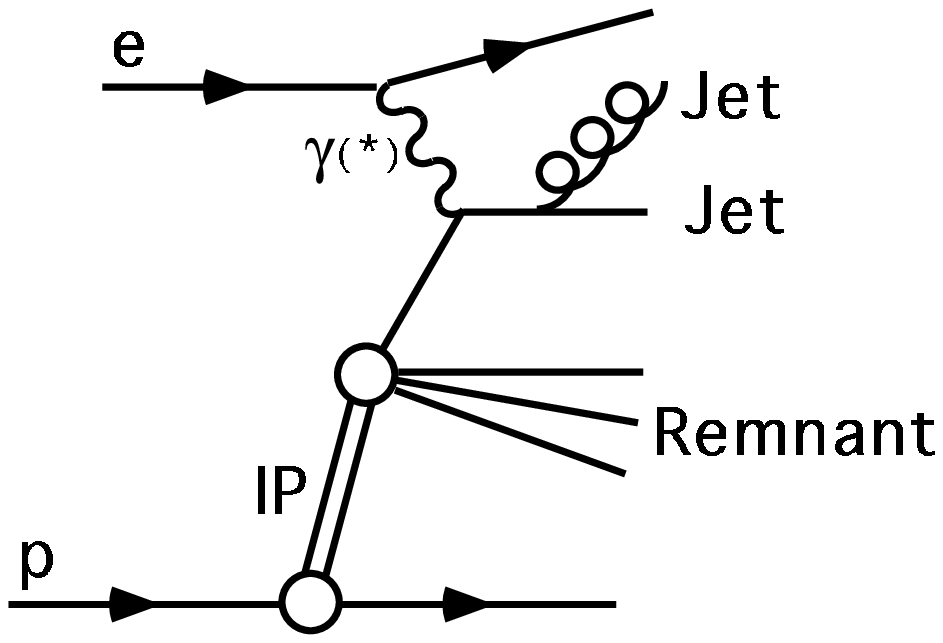,width=0.28\textwidth}}
     \put(0,-5){\bf{(a)}}
     \put(72,-5){\bf{(b)}}
   \end{picture}
 \end{center}
 \scaption {${\cal O} (\alpha_{\rm em} \alpha_{\rm s})$ direct
photon processes leading to diffractive dijet production.
(a) Example QCD-Compton process;
(b) Boson-Gluon Fusion.}
  \label{dijets}
\end{figure}

The electron-pomeron cross section, 
${\rm d} \sigma^{e \pom \rightarrow e X}(\mu, \xgam, \zpom)$, 
is obtained from hard scattering matrix elements 
to leading order in QCD, 
convoluted with parton distributions 
for the pomeron at momentum fraction $\zpom$ and 
for the photon at momentum fraction $\xgam$. 
The scale $\mu$ at 
which the parton distributions
are taken is chosen to be $\pthat$.
In RAPGAP and for direct processes in POMPYT, 
the dominant high $\pthat$ process involving a quark from the exchange is
the QCD-Compton mechanism $\gamma^{(*)} q \rightarrow qg$ 
(figure~\ref{dijets}a), whereas it is
the boson-gluon fusion 
process $\gamma^{(*)} g \rightarrow q \bar{q}$ (figure~\ref{dijets}b)
for a gluon from the exchange.

A set of pomeron parton distributions 
is implemented in the simulations
that is dominated at low scales by 
a gluon distribution peaked at large $\zpom$ \cite{F2D394}.
For the resolved photon component in POMPYT,
the parton distribution functions of the photon are taken from the leading
order GRV parameterisation\cite{GRV}. 
Outgoing charm quarks are 
generated through the photon-gluon and gluon-gluon
fusion processes.
Intrinsic transverse momentum of the incoming partons is simulated for the
photon in POMPYT, but not for the pomeron in either model.

A small contribution from a 
sub-leading exchange ($\reg$) is also included in the simulations, 
as has been found necessary in both photoproduction\cite{H1:gammap}
and DIS\cite{F2D394}. As in the fits to inclusive large rapidity gap data, the
flux factor has the same form as equation (\ref{flux}), with
$\alphareg(t) = 0.50 + 0.90 \ t$ and $b_{_{\reg}} = 2.0 \ {\rm GeV^{-2}}$.
The parton densities for the sub-leading exchange
are taken from a parameterisation for the pion\cite{owens:pion}.

To avoid divergences in the calculation of QCD matrix elements for scattering
from massless partons, a cut is applied at the generator level, 
$\pthatsq > 4 \ {\rm GeV^2}$ (POMPYT) and \linebreak
$\pthatsq > 4.5 \ {\rm GeV^2}$ (RAPGAP). The losses due to these cuts are
negligible for jets with transverse momentum $\ptj > 5 \ {\rm GeV}$.
Higher order effects in the QCD cascade are simulated using  
parton showers \cite{parton:shower} in the leading $\log (\mu)$
approximation (MEPS). For systematic studies, the parton showering is
replaced in RAPGAP by the colour dipole 
model (CDM) as implemented in ARIADNE \cite{ARIADNE}. Hadronisation
is simulated in both models using 
the Lund string model \cite{LUNDstring}.

As explained
in section~\ref{results},
Monte Carlo simulations are also used to
compare the meas-ured hadron level cross sections with the 
predictions of theoretical models.
Several sets of pomeron parton densities are implemented in the RAPGAP and
POMPYT simulations for this purpose. 
Additional models based on the diffractive interaction of the $q \bar{q}$
fluctuation of the photon and on soft colour interactions in inclusive DIS
are also tested.

\section{Experimental procedure}
\label{procedure}

The measurements presented here are based on H1 $e^+ p$ 
data\footnote{The lepton beam particle is referred to in this paper as `the
electron'.}
taken during 1994.
An integrated luminosity of $2.24 \pm 0.03 \ {\rm pb^{-1}}$ is used for
the photoproduction analysis and $1.96 \pm 0.03 \ {\rm pb^{-1}}$ for 
DIS.
Detailed descriptions of the photoproduction and DIS measurements can be 
found in \cite{theseJan} and \cite{theseBertrand} respectively.

\subsection{The H1 detector and kinematic reconstruction}
\label{h1det}

The H1 detector is described in detail elsewhere \cite{h1:detector}. A brief
account of the components that are most relevant to the present analyses is 
given here. The coordinate system convention for the experiment defines the
forward, positive $z$ direction as being that of the outgoing proton beam, 
corresponding to the region where pseudorapidity, \linebreak
$\eta = - \ln \tan \theta / 2$, is positive. 

A finely segmented electromagnetic and hadronic liquid argon calorimeter (LAr)
provides smooth and hermetic coverage in the range 
$-1.5 \lapprox \eta \lapprox 3.4$ with
energy resolution $\sigma(E) / E \simeq 0.11 / \sqrt E$ for electromagnetic
showers and
$\sigma(E) / E \simeq 0.5 / \sqrt E$ for hadrons ($E$ in GeV) as measured
in test beams. The Backward
Electromagnetic lead scintillator Calorimeter (BEMC) covers the region
$-3.4 \lapprox \eta \lapprox -1.4$ 
with resolution $\sigma(E) / E \simeq 0.10 / \sqrt E$.
Beam induced backgrounds
are heavily suppressed using information from the Time of Flight 
scintillator (ToF),
which is situated immediately backward of the BEMC. Charged track
momenta are measured
in the range $-1.5 \lapprox \eta \lapprox 1.5$ 
in the two large concentric drift
chambers (CJC) of the central tracker, located inside a $1.15 \ {\rm T}$ 
solenoidal magnetic 
field. 
Luminosity is measured by detecting electrons and photons from the 
bremsstrahlung process $ep \rightarrow ep \gamma$ in photon and 
electron crystal
calorimeters situated at 
$z = -103 \ {\rm m}$ and $z = -33 \ {\rm m}$ respectively.

In the photoproduction analysis, 
the final state electron is detected in the electron 
tagger of the luminosity system (eTag), which 
has an acceptance such that $Q^2 < 10^{-2} \ {\rm GeV^2}$. 
The measurement of the scattered electron energy $E^{\prime}_e$
is used to reconstruct $y$ according to $y = 1 - E^{\prime}_e / E_e$,
where $E_e = 27.5 \ {\rm GeV}$ is the electron beam energy. 

In DIS, final state electrons at polar scattering angles 
$156^{\circ} < \theta_e < 174^{\circ}$ are
identified using the BEMC in combination with a hit in multi-wire 
proportional chambers (BPC) mounted directly in front of it. 
The inclusive kinematic variables are calculated as
\begin{eqnarray}
Q^2=4 E_e E^{\prime}_e \cos^2 \frac{\theta_e}{2} \hspace{1.5cm} 
y=1-\frac{E^{\prime}_e}{E_e} \sin^2 \frac{\theta_e}{2} \hspace{1.5cm}
x=\frac{Q^2}{s y} \ ,
\end{eqnarray}
using the measurements of $E^{\prime}_e$ in the BEMC and $\theta_e$ from the 
associated BPC hit in
combination with the interaction vertex reconstructed in the central tracker.

The combined use of the Proton Remnant 
Tagger (PRT) scintillator surrounding the forward
beampipe at $z=24 \ {\rm m}$, three 
drift chamber layers of the Forward Muon Detector (FMD) at
$6 \lapprox z \lapprox 7 \ {\rm m}$
and the copper-silicon sandwich
`plug' calorimeter nearest the beam-pipe at $z \simeq 5 \ {\rm m}$
provides sensitivity to hadronic energy flow at pseudorapidities up to 
$\eta \simeq 7.5$ \cite{F2D394,H1:gammap}.

\subsection{Large rapidity gap event selection}

The trigger used to collect the photoproduction data was based on
a scattered electron detected in the eTag and at least one track
in the CJC. 
The DIS data were triggered
on the basis of an energy cluster in the BEMC fulfilling the timing criteria
of the ToF.

A number of selection criteria are applied in order to restrict the 
measurements to regions in which the acceptance
is large and uniform and to suppress backgrounds. 
In both measurements, the reconstructed position of the event
vertex is required to lie within $30 \ {\rm cm}$ $(\sim 3 \sigma)$ 
of the mean interaction point
in the $z$ coordinate.
In the photoproduction analysis, the scattered
electron energy must correspond to the region $0.25 < y < 0.7$. 
To suppress the case in which a bremsstrahlung process is superimposed
on a photoproduction event, the energy measured in the  
photon detector of the luminosity system is required to be less than
$2 \ {\rm GeV}$.

The DIS analysis is restricted to the region
$7.5 < Q^2 < 80 \ {\rm GeV^2}$ and $0.1 < y < 0.7$.
Photoproduction events with high energy hadrons in the BEMC
are removed from the DIS sample by imposing three conditions: 
the energy of the scattered electron 
$E^{\prime}_e$ must be greater than $8 \ {\rm GeV}$, the radius of the 
electromagnetic cluster associated with the electron 
candidate must be less than 
$5 \ {\rm cm}$ and there must be a BPC 
hit with a transverse distance of
less than $5 \ {\rm cm}$ from the centroid of the cluster. 
To suppress events with initial state electromagnetic radiation, 
the summed $E - p_z$ of 
all reconstructed particles including the electron is required to be greater 
than $40 \ {\rm GeV}$.

Events with large forward rapidity gaps are selected in 
both analyses by demanding that there be no recorded signal above 
noise levels in the PRT, the FMD and the plug calorimeter. 
In addition, the most forward calorimeter
cluster with a measured energy of 400 MeV or more must have 
$\eta < 3.2$. These selection criteria ensure that the
forward limit of the system $X$ is contained within the main detector
components and the 
pseudorapidity gap separating $X$ and $Y$ spans at least the 
region $3.2 < \eta \lapprox 7.5$. The range of
accessible masses of the system $X$ is thus extended considerably
compared to previous
diffractive jet analyses at 
HERA\cite{H1:gpdiffjet,ZEUS:gpdiffjet,ZEUS:jetnew}. 
The upper limit of the pseudorapidity gap restricts the
measurements to the region $\my < 1.6 \ {\rm GeV}$ and 
$|t| < 1.0 \ {\rm GeV^2}$.

\subsection{Invariant mass and jet reconstruction}
\label{jetalgo}

The hadronic
system $X$ is detected and measured in the LAr and BEMC calorimeters together
with the CJC.
The mass $\mx$ is
reconstructed by combining 
tracks and calorimeter clusters in energy flow algorithms
that avoid double counting \cite{H1:thrust,H1:gammap}.
The invariant mass of $X$ is then reconstructed according to
\begin{eqnarray}
\mx^2 = (\sum_i E_i)^2 - (\sum_i \vec{p}_i)^2 \ ,
\end{eqnarray}
where the sums extend over all selected tracks and clusters. 
$W^2$ is reconstructed from the measurement of the final state electron
in DIS and using $W^2 = s \ (E - p_z)_{_X} / \ 2 \ E_e$ 
in photoproduction. The remaining diffractive
kinematic variables are computed using
\begin{eqnarray}
\xpom = \frac{\mx^2 + Q^2}{W^2 + Q^2} \hspace{1.5cm}
\beta = \frac{Q^2}{\mx^2 + Q^2} \ ,
\end{eqnarray}
where $t$ and the proton mass squared are neglected.
A cut of
$\xpom < 0.05$ is applied to further reduce non-diffractive contributions.
In the DIS analysis, the data are also restricted to $\xpom > 0.005$ to
remove events in which the system $X$ lies backward of the acceptance region
of the LAr calorimeter.

The large rapidity gap samples are subjected to 
jet searches using a cone algorithm\cite{cone} (radius
$\sqrt{\Delta \eta^2 + \Delta \phi^2} = 1$),
applied to the tracks and clusters included in the reconstruction of $\mx$.
The jet finding takes place in the rest frame of the system $X$ 
(equivalently the $\gamma^{(*)} \pom$ centre of mass frame), with
transverse energies calculated relative to the $\gamma^{(*)}$ axis
in that frame. 
In transforming to the rest frame of $X$, the value of
$|t|$ is fixed at its minimum kinematically allowed value, such that
$p_{_X} = q + \xpom P$.
In the photoproduction case, a simple Lorentz boost in the beam direction 
is thus required. For DIS, the beam 
and $\gamma^*$ axes are not collinear. 

Exactly two jets 
with transverse energy $\etj > 5 \ {\rm GeV}$ are required.
To ensure that the bulk of the jet energy is restricted to the region 
covered by the LAr calorimeter, events are only considered if both jet axes
lie within the region of laboratory pseudorapidity \linebreak
$-1.0 < \eta_{\rm lab}^{\rm jet} < 2.0$ (photoproduction) and
$-1.0 < \eta_{\rm lab}^{\rm jet} < 2.2$ (DIS). After these requirements,
477 events remain in the photoproduction sample and 54 events remain for DIS. 

\begin{center}
\begin{figure}[htbp] \unitlength 1mm
    \begin{picture}(160,47.5)
      \put(-15,-10) {\epsfig{file=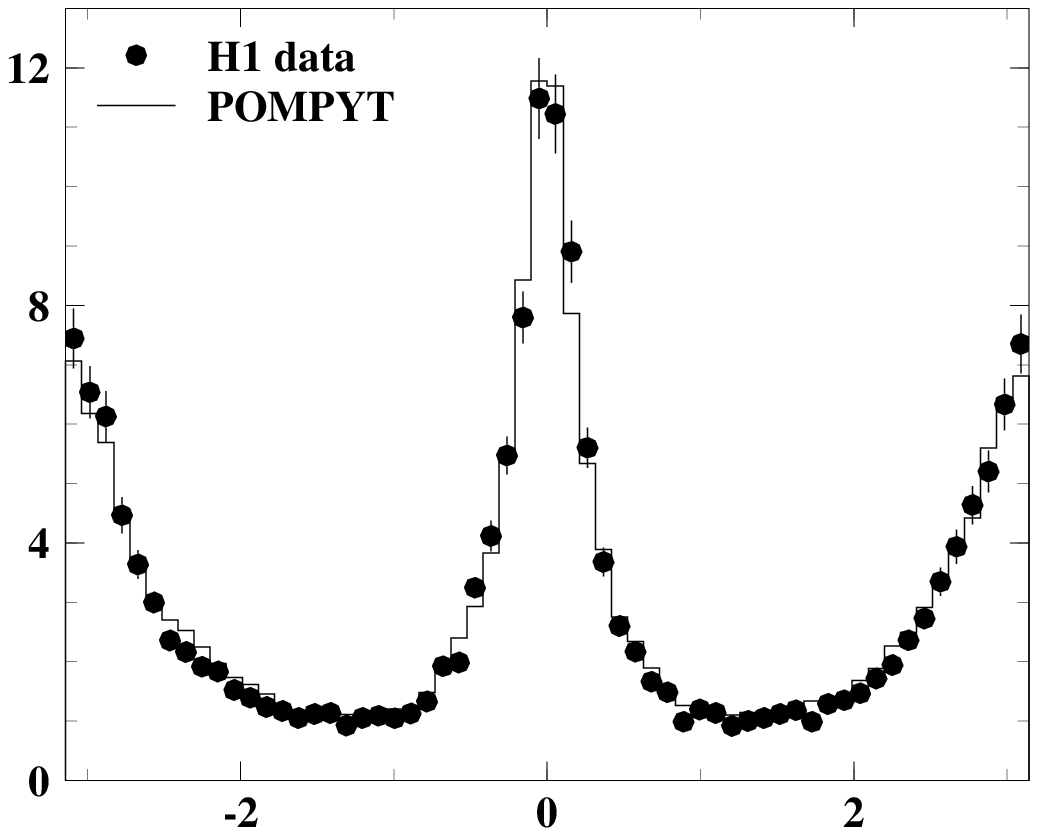,angle=90,width=0.65\textwidth}}
      \put(70,-10) {\epsfig{file=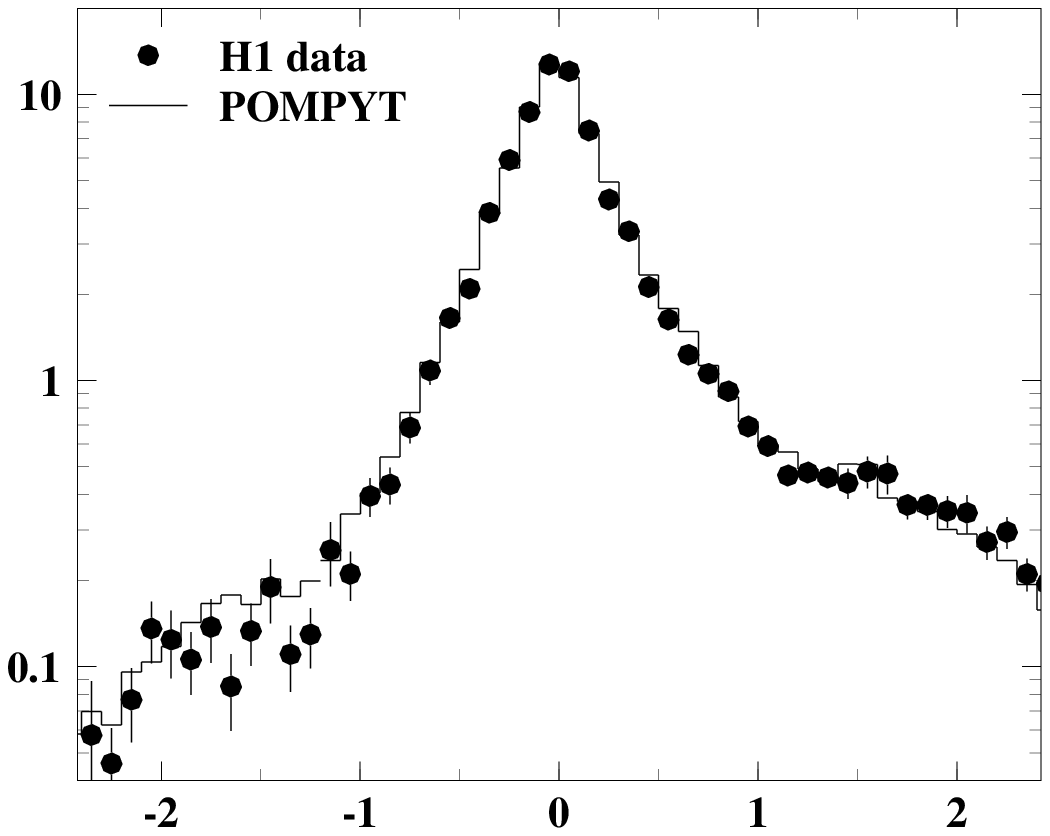,angle=90,width=0.65\textwidth}}
     \put(60,43){\bf{(a)}}
     \put(145,43){\bf{(b)}}
     \put(0,5){\begin{turn}{90}{\boldmath{$\frac{1}{N_{\rm jet}}
\frac{{\rm d} E_{_T}}{{\rm d} \Delta \phi}$} 
(\boldmath{${\rm GeV / radian}$})} \end{turn}}
     \put(85,21.5){\begin{turn}{90}{\boldmath{$\frac{1}{N_{\rm jet}}
\frac{{\rm d} E_{_T}}{{\rm d} \Delta \eta}$} 
(\boldmath{${\rm GeV}$})} \end{turn}}
     \put(47,-5.5){\boldmath $\Delta \phi$ (radians)}
     \put(148,-5.5){\boldmath $\Delta \eta$}
    \end{picture}
  \vspace{-0.25cm}
  \scaption  {The observed distribution
of transverse energy flow about the jet axis for 
the large rapidity gap
photoproduction dijet sample. $\Delta \eta$ and $\Delta \phi$ are the
distances from the jet axis in azimuthal angle and pseudorapidity
respectively. (a) The jet profile in
azimuth integrated over the full range of pseudorapidity. (b) The jet
profile in pseudorapidity integrated over one radian of $\Delta \phi$
about the jet axis.} 
  \label{profiles}
\end{figure}
\vspace{-1.0cm}
\end{center}

The distribution of transverse energy flow about 
the jet axes for the photoproduction measurement is shown in 
figure~\ref{profiles}. For the jet profile in $\Delta \phi$, activity 
throughout the full range of pseudorapity is included in the
plot, which exhibits a clear back-to-back two jet structure. The 
r.m.s. of the distribution in azimuthal difference between the 
jet axes is 
approximately $25^{\circ}$. For the jet profile in $\Delta \eta$, only
transverse energy flow within one radian in azimuth of the jet
axis is shown. The level of activity outside the jet cone is comparable to 
that for inclusive jet photoproduction in the backward 
region \cite{H1:gammapjet1}. It is smaller
in the forward region in the present data, 
due to the reduction in phase space available for underlying activity
implied by the rapidity gap.
The jet profiles in DIS show similar characteristics.

The additional dijet variables $\shatj$, $\xgamj$ and $\zpomj$ are 
reconstructed as specified in equations (\ref{shatetc1} -- \ref{shatetc3}).
The POMPYT and RAPGAP simulations have been used to investigate the 
correlations between the
hadron jets that define the measured cross sections and the underlying
parton dynamics.
The hadronisation process results
in only a small smearing of the jet directions relative to those of the
outgoing partons, with r.m.s. shifts of \linebreak
approximately 0.13 pseudorapidity units and 
$6^{\circ}$ in azimuth. The smearing effects due to hadronisation are
stronger for $\pt$, $\xgam$ and $\zpom$, leading to a resolution of
approximately 20\%, except near
$\xgam = 1$ and $\zpom = 1$,
where events are smeared at the hadron level throughout the regions 
$\xgamj \gapprox 0.6$ and $\zpomj \gapprox 0.6$.

\subsection{Cross section measurement}

The cross sections presented here are defined solely in terms of 
ranges in kinematic variables and jet criteria. They are corrected to the
Born level. 
The kinematic domains to which the cross section measurements 
are corrected are summarised in
table~\ref{table:kinrange}.
The photoproduction cross sections are defined in terms of a
laboratory pseudorapidity region for the jets. In the DIS measurement, the
$y$ and $\xpom$ restrictions imply a similar range of \linebreak
pseudorapidity.
No attempt is made to unfold 
to parton cross sections or to a specific physics process such as 
diffraction, due to the large uncertainties inherent in such a procedure.

\begin{table}[h]
\begin{center}
\begin{tabular}{|c|c|} \hline
PHOTOPRODUCTION & DIS  \\ \hline \hline
$Q^2 < 0.01 \ {\rm GeV^2}$          & $7.5 < Q^2 < 80 \ {\rm GeV^2}$ \\ \hline
$0.25 < y < 0.7$                    & $0.1 < y < 0.7$ \\ \hline
$\xpom < 0.05$                      & $0.005 < \xpom < 0.05$ \\ \hline
\multicolumn{2}{|c|}{$\my < 1.6 \ {\rm GeV}$} \\ \hline
\multicolumn{2}{|c|}{$|t| < 1.0 \ {\rm GeV^2}$} \\ \hline
\multicolumn{2}{|c|}{Exactly two jets with $\ptj > 5 \ {\rm GeV}$} \\ \hline
$-1 < \eta_{\rm lab}^{\rm jet} < 2$ & \\ \hline 
\end{tabular}
\scaption{The kinematic domains in which the photoproduction and DIS 
cross sections for the process $ep \rightarrow eXY$ are measured.}
\label{table:kinrange}
\end{center}
\vspace{-0.5cm}
\end{table}

The data are corrected for detector inefficiencies 
and migrations of kinematic quantities in the reconstruction 
using the POMPYT (photoproduction) and RAPGAP (DIS) Monte Carlo models as
described in section~\ref{mcmods}.
For both measurements, these simulations give an 
adequate description of all relevant reconstructed data distributions.
Migrations about the upper $\xpom$ boundary of 
the measurements
are evaluated with the additional
use of the PYTHIA \cite{PYTHIA} model in photoproduction and the 
DJANGO \cite{DJANGO} model in DIS.
Migrations about the limits of the measurements in $\my$ are studied
using the PHOJET \cite{PHOJET} and PYTHIA models of soft photoproduction and 
the DIFFVM \cite{DIFFVM} simulation of vector meson electroproduction.
Each of these models contains
diffractive events both where the proton remains intact and where it
dissociates.
For the data points presented, the lowest 
overall acceptance is 37\% and the lowest bin 
purity\footnote{Purity is defined as the 
proportion of the 
simulated events reconstructed in an interval that were also 
generated in that interval.} is 32\%
according to the simulations.

Fluctuations in the level of noise in the
forward detector components, resulting in the rejection of
events within the
kinematic range of the measurements, are studied using
a sample of events which were triggered randomly throughout the run
period in which the data were collected.
A correction of $(6.1 \pm 2.0) \%$ is applied.
The expected background from the process $\gamma \gamma \rightarrow q \bar{q}$
has been quantified using the LPAIR Monte Carlo model \cite{lpair}.
Subtractions of
$1.7 \pm 0.3$ events in the photoproduction analysis and $0.7 \pm 0.1$ events
for DIS are made, concentrated at large $\zpomj$ in both cases.
A correction of 1.8\% is applied in the photoproduction measurement to 
account for the loss of signal due to 
the removal of events in which a bremsstrahlung process is overlaid.
Photoproduction background to the DIS measurement has been found to be less
than 1.4\% using the POMPYT simulation.
Beam induced backgrounds are found to be negligible in both 
data samples. 

QED radiative corrections have been shown to be small in previous 
photoproduction measurements \cite{H1:gammapstot}
due to the cut on the energy in the photon detector. 
They are neglected here.
Radiative corrections to the DIS 
measurement are evaluated using the RAPGAP Monte Carlo model interfaced to 
the program HERACLES \cite{HERACLES}. Corrections of up to 10\% 
are applied.

\subsection{Systematic error analysis}
\label{systematics}

The largest contributions to the systematic errors in both measurements
arise from uncertainties in detector calibration. 
\begin{itemize}
\item A 4\% uncertainty in the absolute hadronic energy scale of the LAr 
and a 3\% uncertainty in the fraction of energy carried by tracks 
are reflected in the determination of jet transverse momenta.
Together with a 20\% hadronic energy scale uncertainty of the BEMC,
these uncertainties also affect the measurement of $\xpom$.
The effects on the photoproduction and DIS measurements are different,
mainly because of the different algorithms
used for the selection of tracks and clusters.
In photoproduction, the combined uncertainty 
due to the LAr, BEMC and track scales is approximately 20\% in the cross 
sections differential
in $\eta_{\rm lab}^{\rm jet}$ and $\ptj$ and approximately 15\% for the 
$\xgamj$ and $\zpomj$ distributions. 
These errors arise predominantly from the LAr
and are strongly correlated between data points. For the DIS 
measurement, the LAr and track uncertainties both result in large 
contributions
to the systematic errors. Together with the BEMC uncertainty,
their net result is an uncertainty at the level of 15\% on the measurements, 
which is less strongly correlated between data points.
\item In photoproduction, a 5\% systematic error
arises from the uncertainty in the eTag acceptance averaged over the $y$
range of the measurement. The efficiency of the CJC component of the 
trigger is known to 4.5\% from studies using an independent trigger.
\item In the DIS measurement there is a 1\% uncertainty in the scattered
electron energy $E_e^{\prime}$ and a $1 \ {\rm mrad}$ uncertainty on the
electron scattering angle $\theta_e$. These affect the 
determination of the $\gamma^* \pom$ collision axis and give rise 
to further uncertainties in the cross sections of 6\% and 2\% on average
respectively. 
\item Uncertainties of 20\% in the PRT 
efficiency and 30\% in the plug energy scale
result in normalisation uncertainties of
4\% and 2\% of the measured cross sections respectively.
\item The uncertainty in the luminosity of the data samples is 1.5\%.
\end{itemize}

Additional systematic errors arising from the uncertainties in the 
acceptance and mig-ration corrections are estimated by repeating the 
measurements with variations in the kinematic dependences and other details
of the Monte Carlo models. 
\begin{itemize}
  \item The uncertainties arising from the shapes of the $\ptj$ and $\zpomj$
distributions in the models are studied by changing the simulated 
distributions by amounts that are larger than the final errors
on the measurements presented here. The overall
transverse momentum distributions are 
reweighted by $(1 / \pthat)^{\pm 1}$, resulting in average uncertainties of 
7\% in the photoproduction measurement and 5\% for DIS. The overall momentum 
distribution of the partons in the pomeron is reweighted by $\zpom^{\pm 0.2}$ 
and $(1 - \zpom)^{\pm 0.2}$ for photoproduction, leading to 
an average uncertainty of 1\% in the measured cross sections. 
In the DIS case, where the $\zpom$ 
distribution is less well constrained by the present data, factors of 
$\zpom^{\pm 0.5}$ and $(1 - \zpom)^{\pm 0.5}$ are applied, leading to an 
average uncertainty of 5\%.
The ratio of resolved to direct processes is also varied by 50\% in the 
POMPYT simulation, giving rise to an
uncertainty of 2\% on average in the photoproduction measurement.
  \item The $\xpom$ and $t$ distributions are varied by factors
chosen to result in changes that are larger than the level of
precision determined by more inclusive measurements. Reweighting 
the simulated distributions by 
$(1 / \xpom)^{\pm 0.2}$ \cite{H1:gammap,F2D394} leads to uncertainties of
1\% in the measured cross sections on average. The mean uncertainty from 
varying the $t$ distribution by factors $e^{\pm 2 t}$ \cite{ZEUS:LPS} is 3\%.
  \item The number of events in the Monte Carlo models that migrate into the 
sample from the unmeasured region
$\xpom > 0.05$ is varied by $\pm 50\%$, leading to a mean uncertainty of 
6\% in the photoproduction cross sections and 7\% in DIS. 
  \item A total error of 6\% in the measured 
cross sections accounts for uncertainties in the migrations across the
boundary $\my < 1.6 \ {\rm GeV}$. This error is estimated using the Monte
Carlo models that include proton dissociative processes.
The $\my$ distributions, the ratio of single to
double dissociation cross sections and the fragmentation 
scheme for proton dissociation are varied in the models in the manner described
in \cite{H1:gammap,F2D394}.
  \item Uncertainties due to the modelling of hadronisation are
estimated from the difference between the results obtained in the DIS 
analysis using the colour dipole and the parton showering 
simulations. These
errors are at the level of 3\% of the measured cross sections.
\end{itemize}

In the photoproduction measurement, the dominant source of error is the
hadronic energy scale of the LAr. In DIS, the statistical errors are dominant.

\section{Results and discussion}
\label{results}

In this section, differential cross sections are presented
for dijet production in the photoproduction and DIS
kinematic regions specified in table~\ref{table:kinrange}. 
The results are given
in figures~\ref{resultplot1} --~\ref{resultplot3} and
table~\ref{resulttable}. 
In all figures, the inner error bars show the 
statistical errors and the outer error bars show 
the statistical errors added in quadrature with those systematic 
uncertainties that vary from data point to data point. The
shaded bands show overall normalisation uncertainties, 
which contain contributions from the
LAr, BEMC and tracker scale calibrations in the photoproduction case. 

The data
are compared with models of hard diffraction
using the POMPYT and RAPGAP simulations described in section~\ref{mcmods}.
For the pomeron,
three sets of light quark and gluon distributions are 
implemented in the Monte Carlo generators at a scale $\mu = \pthat$
for comparisons with the measurements. These
correspond to the results of leading
order DGLAP fits to the H1 measurement of 
$F_2^{D(3)}$ (section 6 of \cite{F2D394}). 
The first set of parton distributions (labelled `$F_2^D$ fit 1' 
in the figures)
is obtained from a fit
in which only quarks contribute to the pomeron structure at
the starting scale $\mu_0^2 = 3 \ {\rm GeV^2}$.
This set does not give a good description of $F_2^{D(3)}$ and is used here
only
to illustrate the sensitivity of the measurement to the partonic structure
of the interaction.
In two further
fits, gluons are introduced at the starting scale in addition to quarks.
Both of these fits give an acceptable description of $F_2^{D(3)}$ and in
both cases, the fraction of the pomeron momentum carried by gluons is 
between 
80\% and 90\% for $4.5 < \mu^2 < 75 \ {\rm GeV^2}$.
These two fits
differ in the parameterisation used for the parton densities such that the 
gluon distribution at low scales is either relatively flat in 
$\zpom$ (labelled `$F_2^D$ fit 2')
or is peaked in the region of large $\zpom$ (labelled `$F_2^D$ fit 3'). 
They are hereafter
referred to as the `flat' and `peaked' gluon solutions respectively.

A sub-leading exchange is included for all three sets of parton distributions
in the manner
described in \cite{F2D394}. The resulting predicted contribution is 
between 10\% and 20\% of the
measured cross section in all photoproduction and DIS 
simulations (see figure ~\ref{resultplot1}). 

A measure of the theoretical uncertainties in the predictions of the models
is obtained by varying the details of the simulations.
When the CDM scheme is used in place of parton showering in RAPGAP,
the predicted
dijet rate increases by approximately 25\%. 
Taking
$\mu = \pthat / 2$ or $2 \pthat$ in either POMPYT or RAPGAP alters the
predictions at the level of 20\%. Taking the
LAC-1 \cite{lac1} 
parameterisation of the photon parton 
distributions~\footnote{The LAC-1 parameterisation 
has a larger gluon content than the GRV parameterisation.} 
in POMPYT
in place of GRV changes the
predictions by up to 10\%. 
Interference
between the pomeron and sub-leading exchanges was found to be possible
in \cite{H1:gammap,F2D394}, but is not considered here. Higher order 
corrections to the models beyond those already
simulated by parton showering may also alter the predicted cross sections,
though their effects have been found to be small in inclusive dijet
production \cite{H1:gammapjet2}.

\subsection{Jet rates and transverse momentum dependence}

\begin{figure}[htbp] \unitlength 1mm
 \begin{center}
   \begin{picture}(100,200)
    \put(0,115){\epsfig{file=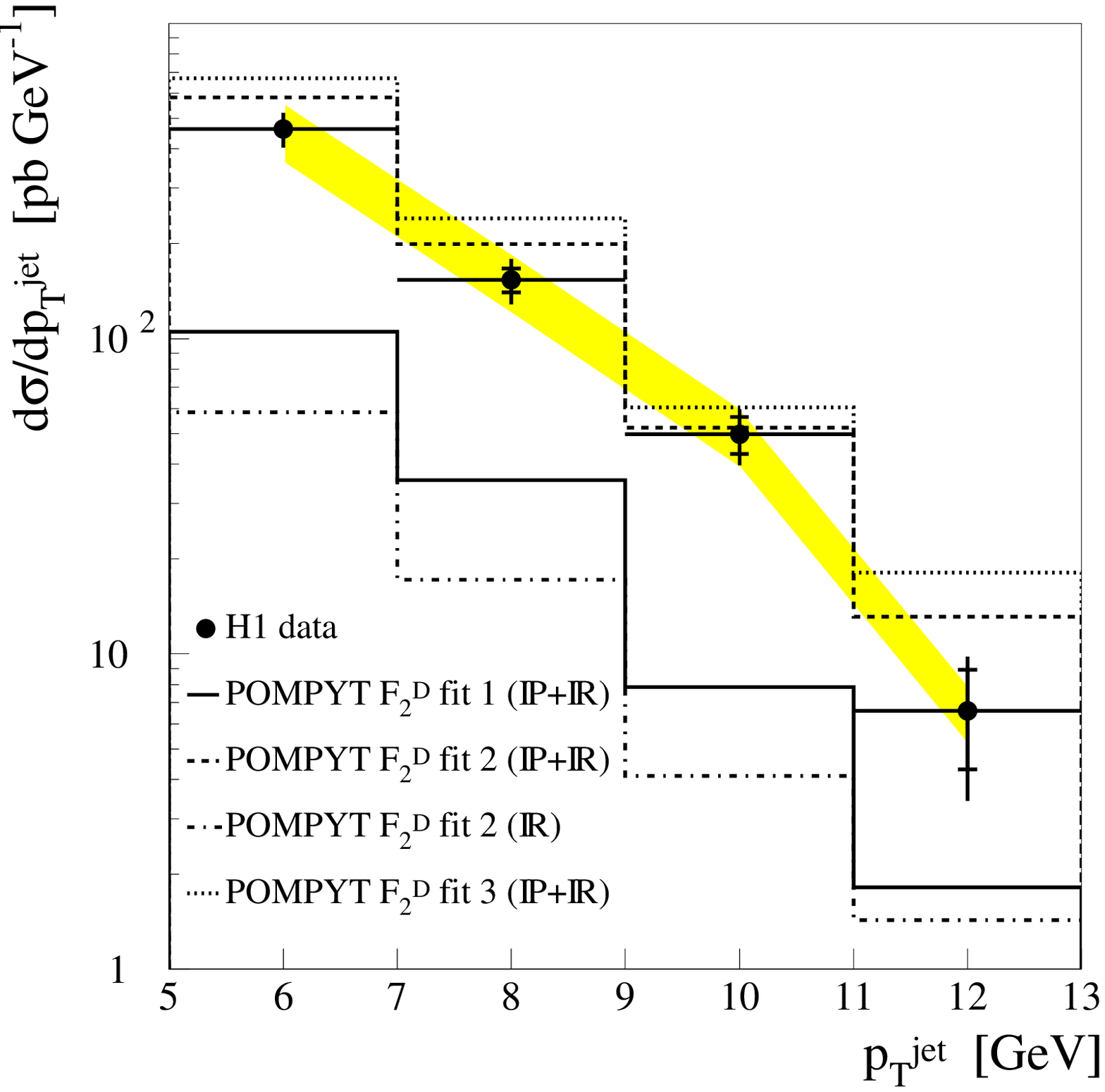,width=0.6\textwidth}}
    \put(0,25){\epsfig{file=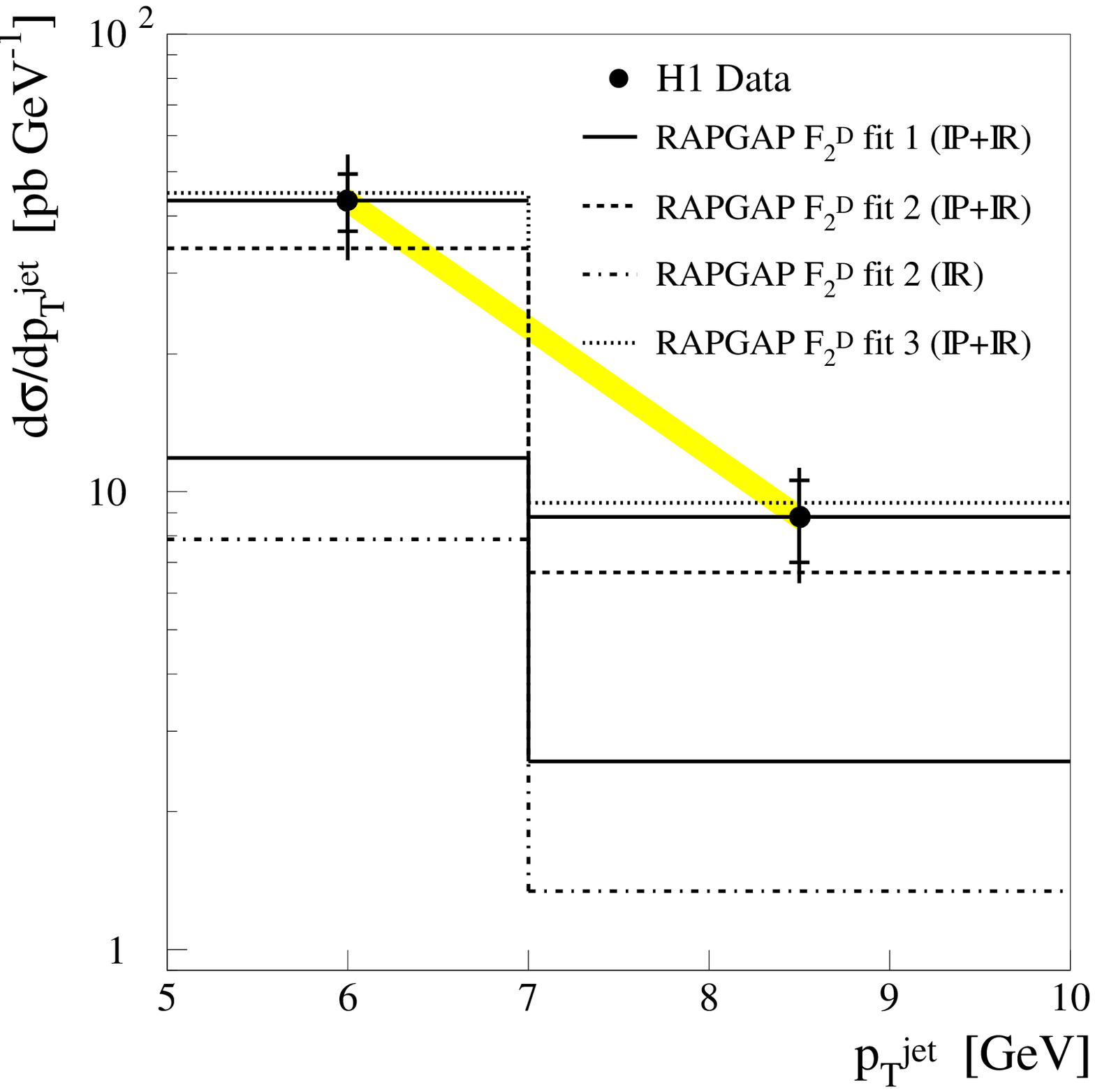,width=0.6\textwidth}}
    \put(90,180){\bf{(a) Photoproduction}}
    \put(90,90){\bf{(b) DIS}}
   \end{picture}
 \end{center}
  \vspace{-3cm}
  \scaption  {Differential cross sections in the component of the jet
momentum transverse to the $\gamma^{(*)} \pom$ collision axis in 
the rest frame of $X$ for the process $ep \rightarrow eXY$ where $X$ contains
two jets. (a) Photoproduction and (b) DIS cross sections measured in the 
kinematic regions
specified in table~\ref{table:kinrange}. There is one entry
in the plots per jet. The shaded bands show the overall normalisation
uncertainties.
The data are compared to the predictions of the
POMPYT (photoproduction) and RAPGAP (DIS) Monte Carlo models with three 
sets of leading order pomeron ($\pom$) and meson ($\reg$) 
parton distributions at a scale set by
$\pthat$: quarks only (labelled $F_2^D$ fit 1), `flat' gluon dominated
(labelled $F_2^D$ fit 2) and `peaked' gluon dominated (labelled 
\protect \linebreak \protect
$F_2^D$ fit 3) parton distributions
at $\pthatsq = 3 \ {\rm GeV^2}$ (see \cite{F2D394}).
The meson component of the `flat' gluon model is 
also shown.}
\label{resultplot1}
\end{figure}

Figures~\ref{resultplot1}a and b show the photoproduction and DIS cross 
sections differential in the jet momentum $\ptj$ transverse to the 
$\gamma^{(*)} \pom$ interaction axis.
In the $\xpom$ region of the present data,
the total mass of the hadronic system containing the jets  
($\mx \lapprox 40 \ {\rm GeV}$) does not greatly exceed the minimum 
accessible value for $\ptj > 5 \ {\rm GeV}$
of $\mx = 10 \ {\rm GeV}$. The $\ptj$
distributions therefore
reflect phase space limitations in addition to dynamics. They are discussed 
here solely in terms of comparisons with the POMPYT and RAPGAP
simulations.

The models in which the diffractive parton distributions consist
solely of quarks at the starting scale underestimate the 
photoproduction and DIS differential cross sections
by factors varying between 3 and 6. The 
models in which the diffractive parton distributions are dominated by hard
gluons are much closer to the data, confirming the conclusions of other
HERA diffractive analyses \cite{F2D393,F2D394,ZEUS:gpdiffjet,ZEUS:jetnew,H1:thrust,H1:eflow,H1:multip,ZEUS:thrust}
that a large gluon component is required in the pomeron parton distributions.
The successful description of the dijet cross sections by the hard gluon
models lends support to the concept of factorisable pomeron parton 
distributions,
appropriate for the modelling of diffractive interactions with 
hard scales other than $Q$.

The sensitivity of 
dijet rates to the partonic composition of the pomeron can be \linebreak
understood
phenomenologically in terms of the hard interactions that can be initiated by
a quark or a gluon. 
If the diffractive parton distributions are quark dominated, then 
the bulk of the diffractive DIS cross section is expected to correspond to
the ${\cal O} (\alpha_{\rm em})$ process $\gamma^* q \rightarrow q$
(figure~\ref{dijets}a without the emission of a gluon in the hard process) 
and the resulting system $X$ must 
be highly aligned with the $\gamma^* \pom$ axis\cite{AJM:ct}. 
The QCD-Compton mechanism (figure~\ref{dijets}a), 
which yields high $\pthat$ outgoing partons, is suppressed
relative to lowest order by ${\cal O} (\alpha_{\rm s})$. By contrast,
for a gluon dominated exchange the lowest order process available is 
boson-gluon fusion (figure~\ref{dijets}b),
yielding two outgoing partons which can have large $\pthat$ due to the
virtuality of the quark propagator. Given that
the overall normalisation of the product
of the pomeron flux and parton distributions is constrained in the
simulations by
the measurement of $F_2^{D(3)}$ \cite{F2D394},
a gluon dominated exchange is thus expected to
result in 
significantly more copious high $\pt$ dijet electroproduction than
a quark dominated exchange. Similar arguments lead to the
same conclusions for direct photoproduction. In resolved photoproduction,
the differences between the predictions for a quark and a gluon dominated
exchange arise from the gluon : quark colour 
factor of $9 : 4$ and other details of 
quark and gluon induced matrix elements.

Comparing the data to the hard gluon simulations in more detail, 
both the `flat' and the `peaked' gluon models reproduce the overall DIS dijet
rate to well within the uncertainties. In photoproduction, the predictions
of both models lie above the data, with the `flat' gluon model closer in
normalisation than the `peaked' gluon
(see also figures~\ref{resultplot2} and~\ref{resultplot3}). 
Assuming that the model based on factorisable evolving diffractive parton
distributions is valid,
the measurements therefore tend to favour the `flat' gluon
solution, though in light of the large
experimental and theoretical uncertainties, no firm conclusions can be
drawn.

\subsection{Photoproduction and rapidity gap survival probability}
\label{gammap}

\begin{figure}[htbp] \unitlength 1mm
 \begin{center}
   \begin{picture}(100,200)
    \put(0,115){\epsfig{file=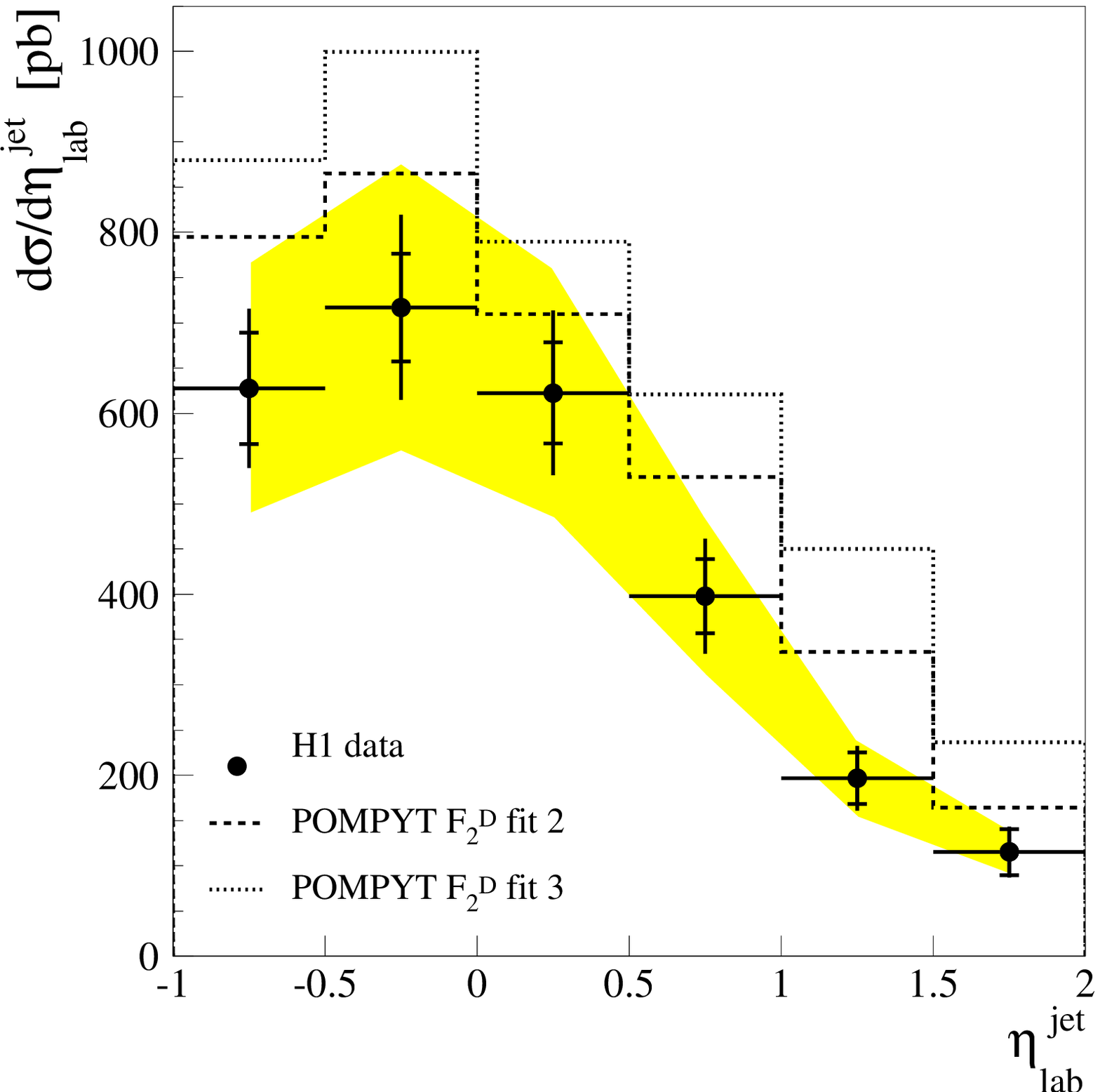,width=0.6\textwidth}}
    \put(0,25){\epsfig{file=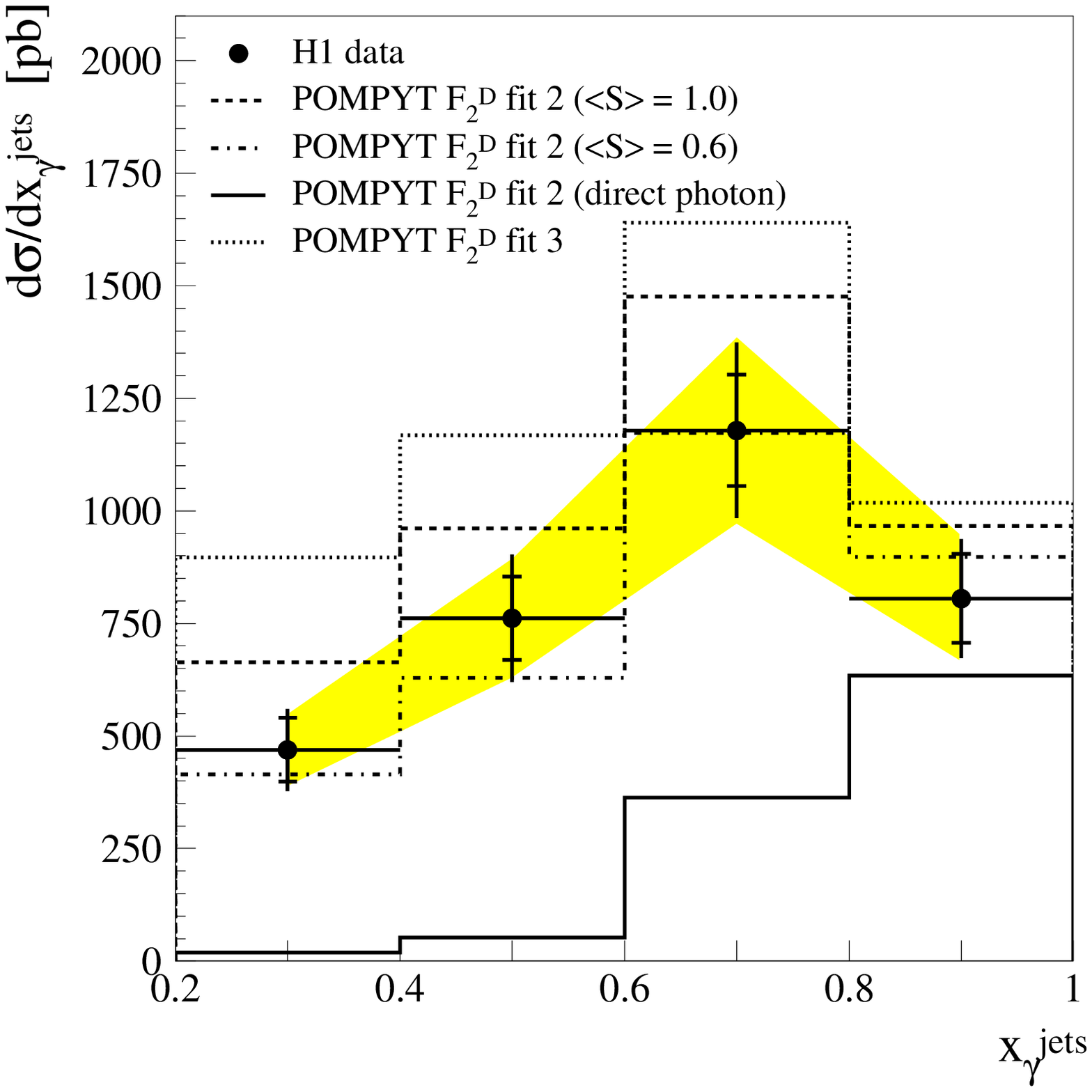,width=0.6\textwidth}}
    \put(90,180){\bf{(a) Photoproduction}}
    \put(90,90){\bf{(b) Photoproduction}}
   \end{picture}
 \end{center}
  \vspace{-3cm}
  \scaption  {Differential cross sections for the production of two jets
in the component $X$ of the process $ep \rightarrow eXY$ in the
photoproduction kinematic region specified in 
table~\ref{table:kinrange}. 
(a) The cross section differential in pseudorapidity in the HERA 
laboratory frame with one entry per jet. (b) The cross section differential
in $\xgamj$ with one entry per event. 
The shaded bands show the overall normalisation
uncertainties.
The data are compared to the predictions of the
POMPYT Monte Carlo model with 
leading order pomeron parton densities at a scale set by
$\pthat$ that are dominated by a `flat'
(labelled $F_2^D$ fit 2) and a `peaked' (labelled 
$F_2^D$ fit 3) gluon distribution
at $\pthatsq = 3 \ {\rm GeV^2}$ (see \cite{F2D394}). In (b),
the predictions for the `flat' gluon model are also shown
with a rapidity gap survival probability of 0.6 applied to events with
$\xgam < 0.8$. The contribution to the model from true direct ($\xgam = 1$)
photon processes is also shown.}
\label{resultplot2}
\end{figure}

The photoproduction cross section differential in jet pseudorapidity in 
the laboratory frame, ${\rm d} \sigma / {\rm d} \eta_{\rm lab}^{\rm jet}$
is shown in figure~\ref{resultplot2}a. The $\eta_{\rm lab}^{\rm jet}$
distribution
is sensitive to the decomposition of the data in 
the variables $\xgamj$ and $\zpomj$, with
the more forward region corresponding broadly to small $\xgam$ and large 
$\zpom$ processes. 
Both hard gluon dominated models acceptably reproduce the shape of the
$\eta_{\rm lab}^{\rm jet}$ distribution. The `flat' gluon solution is the
closer in normalisation.

Figure~\ref{resultplot2}b shows
the photoproduction cross section differential in $\xgamj$.
The contribution in the POMPYT simulation from true
direct photon processes ($\xgam = 1$)
is peaked at the largest $\xgamj$, though there
are significant contributions throughout the region $\xgamj \gapprox 0.6$.
It is clear from the measured distribution that both
direct and resolved photon processes are present in the photoproduction data.

In inclusive jet photoproduction, soft interactions 
between spectator partons in the spatially extended photon and the proton, in 
addition to 
the hard interaction, have been found to be present in
resolved photon processes \cite{H1:incljet}. 
In the diffractive case, multiple 
interactions would be expected to destroy 
rapidity gaps, an effect which has been
parameterised in terms of a 
`survival probability' $S$ \cite{survive}
and which would represent a breaking of diffractive factorisation.
Since no full calculations of
spectator interactions for diffractive photoproduction exist, 
we follow \cite{half:survive} and apply
a constant weighting factor $\langle S \rangle = 0.6$
to all events in
the POMPYT `flat' gluon model that are
generated with $\xgam < 0.8$.
This is a simplistic model, the 60\% survival
probability being an {\it a posteriori} choice. 
Figures~\ref{resultplot2}b 
and~\ref{resultplot3}a show
the effects of the rapidity gap destruction model on the
predicted photoproduction cross sections differential in 
$\xgamj$ and $\zpomj$. The description of the data in both normalisation and 
shape is improved. The photoproduction data are
thus suggestive of the presence of rapidity gap destruction effects, though
the large uncertainties prohibit firm conclusions. This is also true
when the $\zpomj$ distribution is measured separately in 
the two regions $\xgamj < 0.8$ and 
$\xgamj > 0.8$ (not shown).
When rapidity gap destruction effects are
considered for the `peaked' gluon solution, a survival 
probability $\langle S \rangle \sim 0.4$ gives a good description of the
photoproduction data, though spectator interactions cannot explain the 
excess in the predictions of this model
at large $\zpomj$ in DIS (figure~\ref{resultplot3}b).

Studies of diffractive dijet production at the Tevatron 
suggest a breaking of diffractive factorisation \cite{tevatron:jets}. 
From comparisons of the Tevatron data with models based
on pomeron parton distributions extracted from diffractive DIS,
there are indications
that the rapidity gap survival probability may be
as small as 0.1 for $p \bar{p}$ interactions at 
$\sqrt{s} = 1800 \ {\rm GeV}$ \cite{hera:tevatron}.
Within models based on a pomeron with evolving partonic 
structure, there is thus evidence that
any breaking of diffractive factorisation in the present
resolved photoproduction
data is weaker than that in $p \bar{p}$ data
at larger centre of mass energy.
A similar
difference between resolved photoproduction at $W \sim 200 \ {\rm GeV}$ and
$p\bar{p}$ interactions at $\sqrt s  = 1800 \ {\rm GeV}$ is observed in the
fraction of dijet events in which there is a rapidity gap between the 
jets \cite{rapgap:jets}. 
The Tevatron data also suggest that this fraction
is smaller at $\sqrt s = 1800 \ {\rm GeV}$ than at 
$\sqrt s = 630 \ {\rm GeV}$ \cite{tevatron:jets,rapgap:jets}. Data from HERA
and the Tevatron on dijet production in diffractive dissociation and on
rapidity gaps between jets therefore both support the hypothesis that 
rapidity gap survival probabilities decrease with increasing centre of
mass energy \cite{glm:survive}

In \cite{ZEUS:jetnew}, the ZEUS collaboration presented diffractive dijet
photoproduction cross sections measured in a different kinematic region to
that studied here. In particular, the range of $\xpom$ accessed is larger in
the present measurement, resulting in an improved coverage of the low
$\xgamj$ and $\zpomj$ regions.
The 
`flat' gluon model derived from the H1 $F_2^D$ measurements, as implemented
in the POMPYT model, is found to
be compatible with the ZEUS measurements in the specified kinematic
domain.

\subsection{Dependence on fractional momentum from the pomeron}

\begin{figure}[htbp] \unitlength 1mm
 \begin{center}
   \begin{picture}(100,200)
    \put(0,115){\epsfig{file=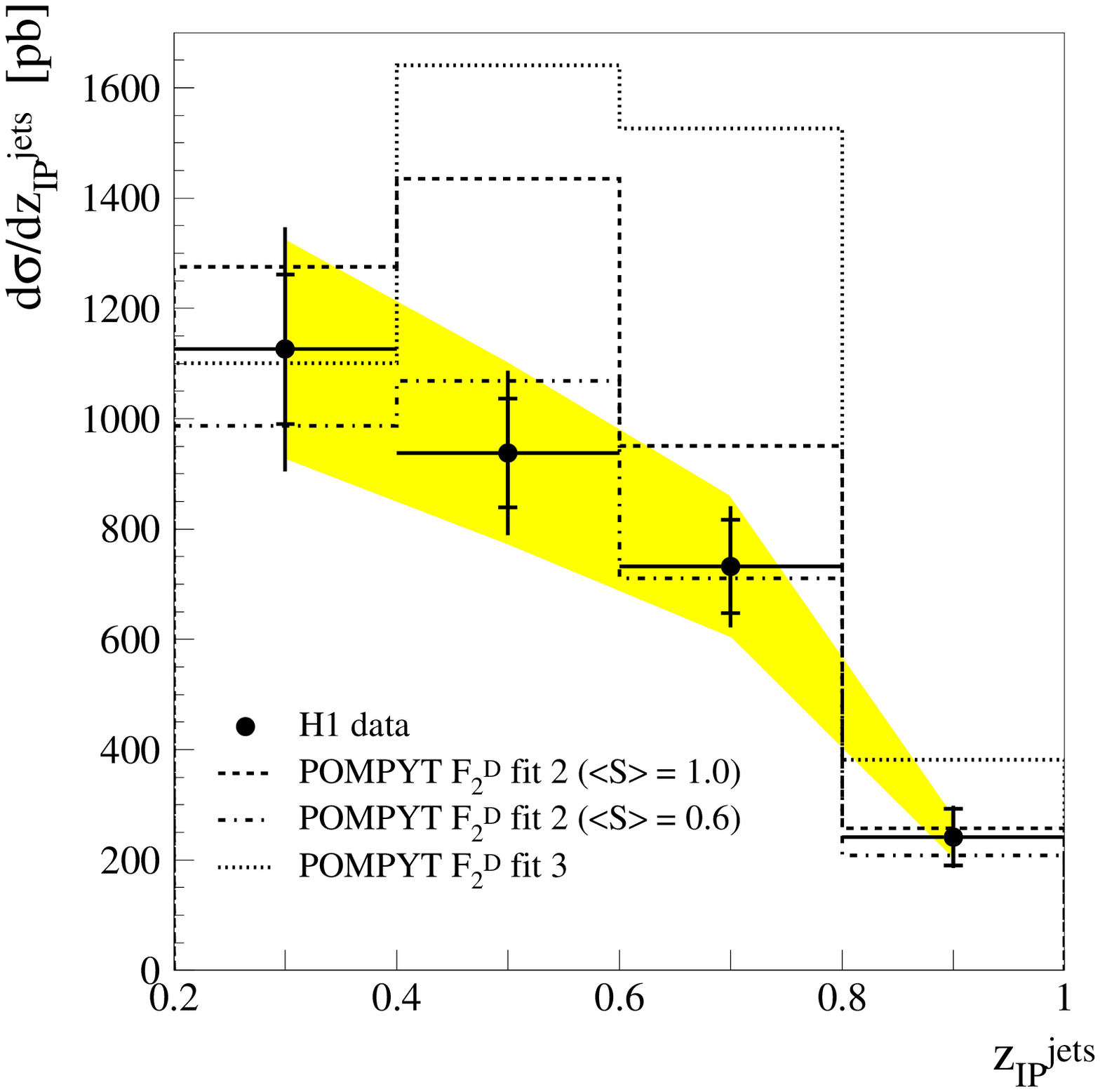,width=0.6\textwidth}}
    \put(0,25){\epsfig{file=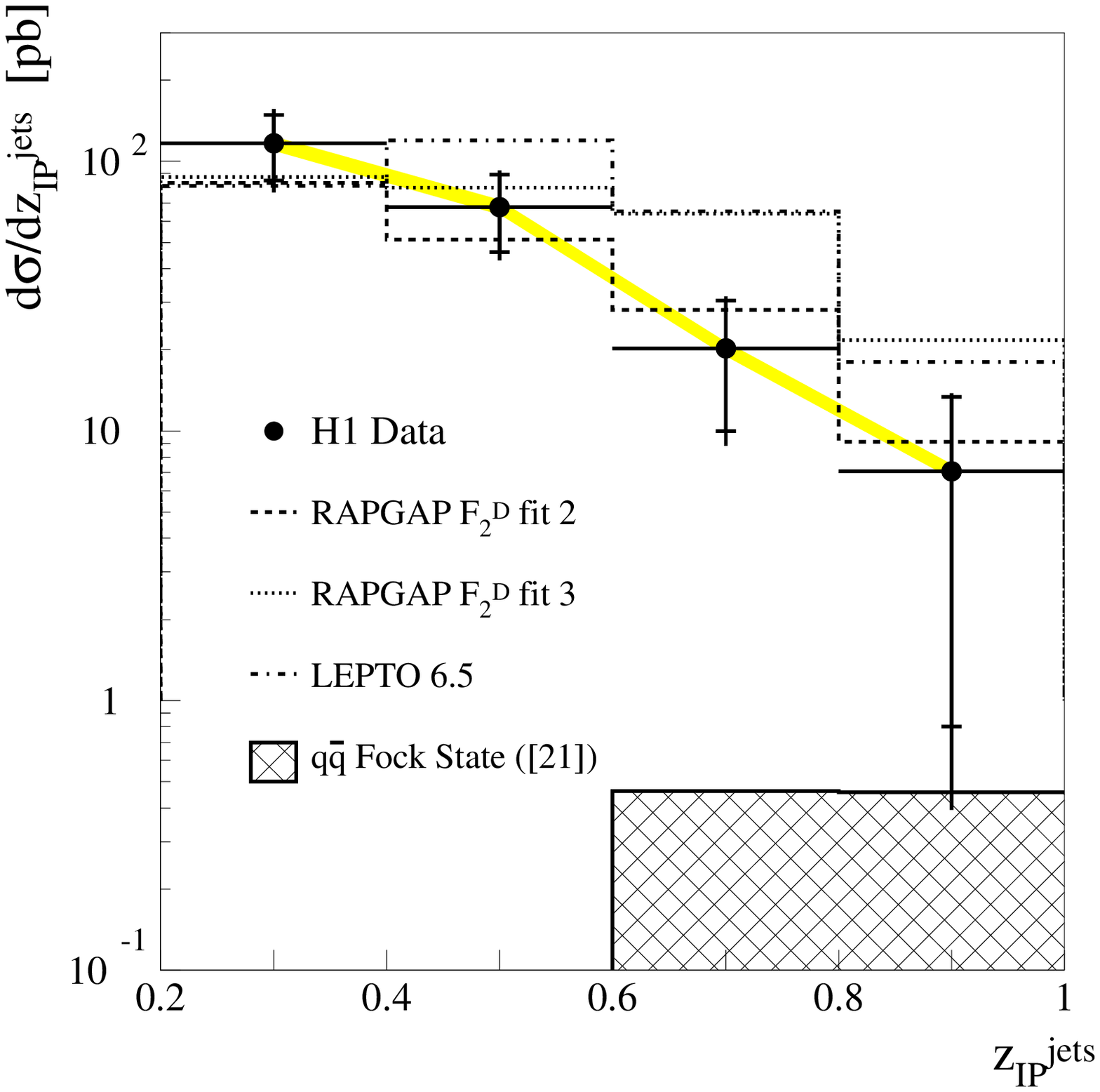,width=0.6\textwidth}}
    \put(90,180){\bf{(a) Photoproduction}}
    \put(90,90){\bf{(b) DIS}}
   \end{picture}
 \end{center}
  \vspace{-3.3cm}
  \scaption  {Differential cross sections in $\zpomj$ for the production of 
two jets in the component $X$ of the process $ep \rightarrow eXY$.
(a) Photoproduction and (b) DIS cross sections measured in the
kinematic regions specified in 
table~\ref{table:kinrange}. 
The shaded bands show the overall normalisation
uncertainties.
The data are compared to the predictions of the
POMPYT (photoproduction) and RAPGAP (DIS) Monte Carlo models with 
leading order parton densities for the pomeron at a scale set by
$\pthat$ that are dominated by a `flat'
(labelled $F_2^D$ fit 2) and a `peaked' (labelled 
$F_2^D$ fit 3) gluon distribution
at $\pthatsq = 3 \ {\rm GeV^2}$ (see \cite{F2D394}).
In (a), the prediction of POMPYT for the `flat' gluon is also shown
with a rapidity gap survival probability of 0.6 applied to events with
$\xgam < 0.8$. Also shown in (b) are the RAPGAP implementation of a
calculation \cite{gg:rapgap} of the 
diffractive scattering of the $q \bar{q}$ fluctuation of the photon and
the LEPTO 6.5 model with a probability of 0.5 for soft colour interactions
to take place.}
\label{resultplot3}
\end{figure}

The photoproduction and DIS dijet production cross sections 
differential in $\zpomj$ are presented
and compared to Monte Carlo predictions in figures~\ref{resultplot3}a 
and ~\ref{resultplot3}b respectively. 
In both kinematic regions, 
there are
significant contributions at the largest $\zpomj$ and
the differential cross sections
increase as $\zpomj$ decreases. 

The calculation 
by Bartels et al. \cite{gg:rapgap}
for the $q \bar{q}$ Fock state arising from transversely polarised photons
is compared to the DIS data via the RAPGAP 
simulation in figure~\ref{resultplot3}b. 
In the picture based on
partonic fluctuations of the photon, 
this contribution is expected to be
the dominant feature of diffractive DIS only at small $\pthat$ and for
$\beta$ values of around $0.5$, larger than those typical of the present
sample \cite{bartels:fullmodel}. 
The diffractive interaction is modelled by coupling two gluons in a net
colour singlet configuration to the outgoing quarks in all
possible combinations. 
The calculation is performed in the double logarithmic
approximation at $t = 0$, such that the cross section is closely related to 
the squared gluon distribution of the proton with momentum fraction 
$x = \xpom$
at a squared scale $\pthatsq \cdot (Q^2 + \mx^2) / \mx^2$. 
The $t$ dependence is modelled
using a parameterisation of the proton form 
factor \cite{dl:form}. 
At the largest $\zpomj$, where the full momentum of the system $X$ is 
carried by the two jets, 
the prediction for the $q \bar{q}$ fluctuation alone
is compatible
with the data, given the large experimental uncertainties. 
At smaller $\zpomj$, this model
falls well short of the data.
Fock states with higher parton multiplicities than
$q \bar{q}$ are presumably dominant in the kinematic regime studied
here, as has also been shown at large $\mx$ and $\pt$ in previous
hadronic final state analyses \cite{H1:thrust,H1:eflow,H1:multip}.
No direct comparisons have been made with models of higher multiplicity
photon fluctuations, though the scattering of the $q \bar{q} g$ Fock state 
where the gluon is the low transverse momentum parton has been identified 
with the 
boson-gluon fusion hard process \cite{hebecker}. 

Certain classes of non-factorising processes which may be present in
resolved photoproduction are predicted to yield a `super-hard' 
contribution to
diffraction for which $\zpom \equiv 1$ \cite{fac:break}. There
is some experimental evidence for such a contribution
from the UA8 collaboration \cite{UA8:super}, at the level of
30\% of the data.
The RAPGAP $q \bar{q}$ Fock state prediction ($\zpom \equiv 1$)
illustrates that a `super-hard' contribution is expected 
to appear in the jet cross sections as a broad
distribution with $\zpomj \gapprox 0.6$. 
This shape contrasts with that observed in
both the photoproduction 
and the DIS data, where there are large 
contributions at $\zpomj \lapprox 0.6$.
`Super-hard' 
diffractive dijet production is therefore not the dominant feature
of the present data, though it is expected to 
become increasingly visible as $|t|$ 
increases \cite{fac:break}. The UA8 measurements were 
for $1 \lapprox |t| \lapprox 2 \ {\rm GeV^2}$, 
whereas $|t| < 1 \ {\rm GeV^2}$ for the data presented here.

The DIS data are also compared with the LEPTO 6.5 \cite{lepto}
model of inclusive DIS in figure~\ref{resultplot3}b. 
In this model, leading order hard 
interactions are convoluted with parton distributions for the proton,
boson-gluon fusion being the dominant process in the low $x$ region
corresponding to the diffractive data.
Soft interactions between the outgoing partons alter the final 
state colour connections without affecting the parton momenta, hence giving
rise to large rapidity gaps \cite{sci}. 
The LEPTO model is able to reproduce the
overall dijet production rate when the probability for soft colour
interactions is around 0.5. The description of the shapes of the
$\ptj$ and $\zpomj$ distributions is similar in quality to that of the
RAPGAP `peaked' gluon model. 


\section{Summary}

Cross sections have been measured for the production of two jets as  
components of the dissociating photon system $X$ in the process 
$ep \rightarrow eXY$ ($\my < 1.6 \ {\rm GeV}$, $|t| < 1 \ {\rm GeV^2}$).
A cone algorithm was used in the rest frame of $X$, requiring
$\ptj > 5 \ {\rm GeV}$ relative to the photon direction in that frame.
Photoproduction cross sections have been measured differentially
in
$\ptj$, $\eta_{\rm lab}^{\rm jet}$, $\xgamj$ and $\zpomj$, with clear
evidence for the presence of resolved as well as direct photon processes.
Diffractive
dijet production has also been studied in DIS, with
cross sections measured differentially in $\ptj$ and $\zpomj$ for the
region $7.5 < Q^2 < 80 \ {\rm GeV^2}$.

The measured dijet
production rates and kinematic distributions
have been compared to models
of inelastic $e \pom$ scattering in
which diffractive parton densities,
extracted from a measurement of inclusive diffractive DIS, are 
evolved using the DGLAP equations to the scale $\pthat$.
Since $\pthat$ rather than $Q$ was chosen for the
scale of the hard interaction here, with $\pthatsq \gg Q^2$
for most of the data, this represents a largely independent 
investigation of the  
validity of
diffractive parton distributions for the description of both 
deep-inelastic and photoproduction interactions.

The dijet measurements can be described by models containing
diffractive parton distributions that are dominated
by hard gluons at low scales. For comparison,
quark dominated diffractive parton densities 
result in dijet rates that are significantly smaller
than those measured. 
Parton distributions in which the pomeron
gluon structure is 
relatively flat as a function of $\zpom$ at low scales
(fit 2 of \cite{F2D394})
describe the data better than those in which the gluon 
distribution is peaked at large $\zpom$ (fit 3 of \cite{F2D394}).
The best description of the combined DIS and photoproduction data is obtained
when a rapidity gap survival probability of 0.6 is applied to the `flat' 
gluon model to account for spectator interactions where there is a photon 
remnant system. However, a survival probability of unity, corresponding to 
no breaking of diffractive factorisation, cannot be excluded.
The photoproduction data are also
compatible with the `peaked' gluon solution when the survival probability
is around 0.4.
The rapidity gap survival probability for resolved photoproduction at HERA 
is thus larger than that for $p \bar{p}$ 
interactions at higher energies at the Tevatron.
There is no evidence for a large `super-hard' pomeron contribution
for which $\zpom \equiv 1$.

When considered in terms of the diffractive scattering of partonic
fluctuations of the virtual photon, the measured dijet rates in DIS 
cannot be described by a simulation of the $q \bar{q}$
Fock state alone. Fluctuations to states containing one or more gluons
are therefore expected to be
dominant in the high $\pthat$ and high $\mx$ region 
investigated here. 

The soft colour interaction mechanism, as implemented in
the LEPTO Monte Carlo model, gives an acceptable description of the DIS data
when the probability for rearrangements in the colour connections
between outgoing partons
is in the region of 0.5.

It has now been shown from inclusive diffractive cross section measurements,
charged particle distributions and multiplicities, energy flow, event shapes
and dijet cross sections that HERA diffractive data are consistently
described by models that assume a $t$ channel exchange with gluon dominated
parton distributions, evolving with the scale of the hard interaction.

\section*{Acknowledgements}

We are grateful to the HERA machine group whose outstanding efforts have 
made and continue to make this experiment possible. We thank the engineers 
and technicians for their work constructing and now maintaining the H1 
detector, our funding agencies for financial support, the DESY technical 
staff for continual assistance and the DESY directorate for the hospitality
which they extend to the non-DESY members of the collaboration. 
We have benefited from interesting discussions with J.\ Bartels and R.\ Engel. 

%


\newpage

\begin{table}[p]
\begin{center}
\begin{scriptsize}
{\large {\bf (a)}} \hspace{0.5cm}
{\begin{tabular}{|p{1.5cm}||p{1.4cm}|p{1.4cm}|p{1.8cm}|p{1.8cm}|} \hline
$\ptj$ & ${\rm d} \sigma / {\rm d} \ptj$ & Stat. error & 
\multicolumn{2}{c|}{Syst. error \ (${\rm pb \ GeV^{-1}}$)} \\ \cline{4-5}
(${\rm GeV}$) & (${\rm pb \ GeV^{-1}}$) & (${\rm pb \ GeV^{-1}}$) & 
Uncorrelated & Correlated \\ \hline
5 - 7   & 462. & 25. & 52. & 103. \\ \hline
7 - 9   & 153. & 13. & 21. &  34. \\ \hline
9 - 11  &  49.7 &  6.7 &  7.4 &  11.1 \\ \hline
11 - 13 &   6.65 &  2.30 &  2.17 & 1.49   \\ \hline
\end{tabular}}
\end{scriptsize}
\end{center}
\end{table}
\begin{table}[p]
\vspace{-0.5cm}
\begin{center}
\begin{scriptsize}
{\large {\bf (b)}} \hspace{0.5cm}
{\begin{tabular}{|p{1.5cm}||p{1.4cm}|p{1.4cm}|p{1.8cm}|p{1.8cm}|} \hline
$\ptj$ & ${\rm d} \sigma / {\rm d} \ptj$ & Stat. error & 
\multicolumn{2}{c|}{Syst. error \ (${\rm pb \ GeV^{-1}}$)} \\ \cline{4-5}
(${\rm GeV}$) & (${\rm pb \ GeV^{-1}}$) & (${\rm pb \ GeV^{-1}}$) & 
Uncorrelated & Correlated \\ \hline
5 - 7   & 43.3 & 6.2 &  9.4 & 3.4 \\ \hline
7 - 10  &  8.82 & 1.78 &  1.65 & 0.71 \\ \hline
\end{tabular}}
\end{scriptsize}
\end{center}
\end{table}
\begin{table}[p]
\vspace{-0.5cm}
\begin{center}
\begin{scriptsize}
{\large {\bf (c)}} \hspace{0.5cm} 
{\begin{tabular}{|p{1.5cm}||p{1.4cm}|p{1.4cm}|p{1.8cm}|p{1.8cm}|} \hline
$\eta_{\rm lab}^{\rm jet}$ & 
${\rm d} \sigma / {\rm d} \eta_{\rm lab}^{\rm jet}$ & Stat. error & 
\multicolumn{2}{c|}{Syst. error \ (${\rm pb}$)} \\ \cline{4-5}
& (${\rm pb}$) & (${\rm pb}$) & 
Uncorrelated & Correlated \\ \hline
-1.0 - -0.5 &  628. & 61. & 64. & 140. \\ \hline
-0.5 - 0.0  &  717. & 60. & 83. & 160. \\ \hline
0.0 - 0.5   &  623. & 56. & 72. & 139. \\ \hline
0.5 - 1.0   &  398. & 41. & 49. &  89. \\ \hline
1.0 - 1.5   &  197. & 29. & 21. &  44. \\ \hline
1.5 - 2.0   &  115. & 26. & 11. &  26. \\ \hline
\end{tabular}}
\end{scriptsize}
\end{center}
\end{table}
\begin{table}[p]
\vspace{-0.5cm}
\begin{center}
\begin{scriptsize}
{\large {\bf (d)}} \hspace{0.5cm} 
{\begin{tabular}{|p{1.5cm}||p{1.4cm}|p{1.4cm}|p{1.8cm}|p{1.8cm}|} \hline
$\xgamj$ & ${\rm d} \sigma / {\rm d} \xgamj$ & Stat. error & 
\multicolumn{2}{c|}{Syst. error \ (${\rm pb}$)} \\ \cline{4-5}
& (${\rm pb}$) & (${\rm pb}$) & 
Uncorrelated & Correlated \\ \hline
0.2 - 0.4 &  469. &  71. &  57. &  85. \\ \hline
0.4 - 0.6 &  762. &  93. & 106. & 137. \\ \hline
0.6 - 0.8 & 1180. & 120. & 150. & 210. \\ \hline
0.8 - 1.0 &  806. &  98. &  89. & 145. \\ \hline
\end{tabular}}
\end{scriptsize}
\end{center}
\end{table}
\begin{table}[p]
\vspace{-0.5cm}
\begin{center}
\begin{scriptsize}
{\large {\bf (e)}} \hspace{0.5cm} 
{\begin{tabular}{|p{1.5cm}||p{1.4cm}|p{1.4cm}|p{1.8cm}|p{1.8cm}|} \hline
$\zpomj$ & ${\rm d} \sigma / {\rm d} \zpomj$ & Stat. error & 
\multicolumn{2}{c|}{Syst. error \ (${\rm pb}$)} \\ \cline{4-5}
& (${\rm pb}$) & (${\rm pb}$) & 
Uncorrelated & Correlated \\ \hline
0.2 - 0.4 & 1130. & 140. & 170. & 200. \\ \hline
0.4 - 0.6 &  938. &  99. & 112. & 169. \\ \hline
0.6 - 0.8 &  732. &  85. &  70. & 132. \\ \hline
0.8 - 1.0 &  242. &  52. &  23. &  44. \\ \hline
\end{tabular}}
\end{scriptsize}
\end{center}
\end{table}
\begin{table}[p]
\vspace{-0.5cm}
\begin{center}
\begin{scriptsize}
{\large {\bf (f)}} \hspace{0.5cm} 
{\begin{tabular}{|p{1.5cm}||p{1.4cm}|p{1.4cm}|p{1.8cm}|p{1.8cm}|} \hline
$\zpomj$ & ${\rm d} \sigma / {\rm d} \zpomj$ & Stat. error & 
\multicolumn{2}{c|}{Syst. error \ (${\rm pb}$)} \\ \cline{4-5}
& (${\rm pb}$) & (${\rm pb}$) & 
Uncorrelated & Correlated \\ \hline
0.2 - 0.4   & 117. & 32. & 25. & 9. \\ \hline
0.4 - 0.6   &  67.7 & 21.7 & 11.8 & 5.4 \\ \hline
0.6 - 0.8   &  20.2 & 10.2 &  5.1 & 1.6 \\ \hline
0.8 - 1.0   &   7.09 &  6.31 &  2.33 & 0.57 \\ \hline
\end{tabular}}
\end{scriptsize}
\scaption{Tables summarising the data points shown in 
figures~\ref{resultplot1} --~\ref{resultplot3} for the kinematic
regions specified in table~\ref{table:kinrange}. Differential cross sections,
statistical errors, systematic errors that vary from data point to data
point and overall normalisation uncertainties are given. (a,b) The 
photoproduction (a) and \protect \linebreak \protect
DIS (b) cross sections differential in the jet
momentum transverse to the $\gamma^{(*)} \pom$ interaction axis in the
rest frame of $X$. (c) The photoproduction
cross section differential in laboratory pseudorapidity. 
(d) The photoproduction
cross section differential in $\xgamj$. 
(e,f) The photoproduction (e) and DIS (f) cross sections differential in 
$\zpomj$. There is one entry in the cross section per jet in (a -- c) and
one entry per event in (d -- f).}
\label{resulttable}
\end{center}
\end{table}

\end{document}